\documentclass[twocolumn]{aa}
\usepackage{psfig}
\usepackage{txfonts}
\def\deg{$^{\circ}$}

\def\extinct{A$_{\rm V}$}
\def\feh{[Fe/H]}
\def\logg{$\log g$}
\def\teff{${\rm T}_{\rm eff}$}
\def\logteff{$\log{\rm T}_{\rm eff}$}

\def\cw{$c$}
\def\hwidth{$b$}
\def\itime{$t$}
\def\itimemin{\itime$_{\rm min}$}
\def\itimemax{\itime$_{\rm max}$}
\def\sigcw{$\sigma_c$}
\def\sighwidth{$\sigma_b$}
\def\sigitime{$\sigma_t$}
\def\Pmut{P$_{\rm m}$}
\def\Precomb{P$_{\rm r}$}
\begin{document}

\title{Evolutionary design of photometric systems\\and its application
to Gaia}

\author{C.A.L.\ Bailer-Jones\inst{1}\fnmsep\inst{2}\fnmsep\thanks{Emmy Noether Fellow of the Deutsche Forschungsgemeinschaft}}

\institute{Carnegie Mellon University, Department of Physics, 5000
Forbes Ave., Pittsburgh, PA 15213, USA
\and Max-Planck-Institut f\"ur Astronomie, K\"onigstuhl 17, 69117
Heidelberg, Germany\\
\email{calj@mpia-hd.mpg.de}
}

\date{Submitted 1 December 2003; Accepted 19 February 2004} 


\abstract{

How do I find the optimal photometric system for a survey?  Designing
a photometric system to best fulfil a set of scientific goals is a
complex task, demanding a compromise between often conflicting
scientific requirements, and being subject to various instrumental
constraints.  A specific example is the determination of stellar
astrophysical parameters (APs) -- effective temperature, surface
gravity, metallicity etc.\ -- across a wide range of stellar types.  I
present a novel approach to this problem which makes minimal
assumptions about the required filter system.  By considering a filter
system as a set of free parameters (central wavelengths, profile
widths etc.), it may be designed by optimizing some figure-of-merit
(FoM) with respect to these parameters.  In the example considered,
the FoM is a measure of how well the filter system can `separate'
stars with different APs.  This separation is vectorial in nature, in
the sense that the local directions of AP variance are preferably
mutually orthogonal to avoid AP degeneracy.  The optimization is
carried out with an evolutionary algorithm, a population-based
approach which uses principles of evolutionary biology to efficiently
search the parameter space.
This model, HFD (Heuristic Filter Design), is applied to the design of
photometric systems for the Gaia space astrometry mission. The
optimized systems show a number of interesting features, not least the
persistence of broad, overlapping filters. These HFD systems perform
as least as well as other proposed systems for Gaia -- as
measured by this FoM -- although inadequacies in all of these systems
at removing degeneracies remain. Ideas for improving the model are
discussed.  The principles underlying HFD are quite generic and may be
applied to filter design for numerous other projects, such as the
search for specific types of objects or photometric redshift
determination.


}

\maketitle

\section{Introduction}\label{intro}

Surveys of stellar populations are directed at improving
our understanding of their formation and evolution. One of the most
important ingredients of such surveys is stellar photometry and/or
spectroscopy as a means to determine fundamental stellar
parameters. These are, in the first instance, atmospheric parameters --
effective temperature, surface gravity and chemical abundances -- from
which we may derive stellar masses, radii and ages.

A fundamental question facing the designers of such surveys is what
spectra and/or photometric systems are optimal for this purpose.
Given the constraints of telescope size and survey duration, the
designer must generally trade off spectral resolution with
signal-to-noise ratio (SNR), limiting magnitude, number of sources and
sky coverage. Some aspects of the design may be obvious from the
scientific goals, but the optimal settings of many others will remain
uncertain.

Deep surveys of large numbers of objects will often be forced to
employ photometry rather than spectroscopy due to confusion and SNR
considerations. Given well defined scientific goals, the designer must
decide how many filters to use, with what kind of profiles, where to
locate them in the spectrum and how much integration time to assign to
each. This is usually achieved via a manual inspection of typical
target spectra.   But if the survey is intended to establish multiple
astrophysical parameters (APs) across a large and varied population of
objects, then this method of filter design is unlikely to be very
efficient or even successful. Manually placing numerous filters and
adjusting their profiles to simultaneously satisfy many different --
and often conflicting -- requirements is likely to be extremely
difficult, especially given the vast number of permutations of filter
parameters possible.   Even if a reasonable filter system could be
constructed in this way, we would not know whether a better filter
system exists subject to the same constraints. Is there not a more
systematic approach to constructing filter systems?

The approach developed in this article is to use a representative grid
of (synthetic) spectra with known APs to construct a filter system in
a heuristic fashion.  The grid represents the scientific goals of the
survey.  Let us assume that for a given filter system we can calculate
a figure-of-merit which is a measure of how accurately the filter
system can determine the APs of the grid spectra.  If we consider the
filter system as a set of free parameters (central wavelengths,
widths, profile shapes etc.), then we may construct a filter system by
optimizing the figure-of-merit with respect to these filter
parameters. This approach has the advantage that it can exploit
the extensive literature on optimization techniques.  The specific
technique used here is a type of evolutionary algorithm, a
population-based technique designed to perform a stochastic yet
directed search of the parameter space, adopting features of
biological evolution (section~\ref{EAs}).

The underlying principle of my approach is to make few prior
assumptions about the required filter system and to let
the optimization proceed freely within the constraints
laid down by the scientific goals and other instrumental
considerations.

The model itself, HFD (Heuristic Filter Design), will be described in
detail in section \ref{hfd_model}, but it is worth highlighting now
that a crucial aspect is to establish a suitable figure-of-merit. The
most obvious would be some average (over the grid) of the precision
with which APs are determined. This could be achieved by any one of
several regression methods -- e.g.\ nearest neighbours or neural
networks -- used to approximate the mapping between the data space and
the AP space, although doing this well for the multiparameter stellar
problem is far from trivial (Bailer-Jones \cite{bj02},
\cite{bj03}). Furthermore, because HFD works through many (ca.\ 10$^5$)
candidate filter systems, fitting a high-dimensional regression model
in each case would be unbearably time consuming.

It turns out that an explicit determination of the performance of the
filter system in these terms is not actually necessary.  A suitable
figure-of-merit can be constructed when we consider what a filter
system does.  Its primary function is to define metrics (e.g.\
colours) which cluster together similar objects and which separate out
dissimilar objects.  A simple example is star--quasar separation. If
the filter system is designed to determine a continuous AP it should
separate objects in proportion to their differences in this AP. In
doing this it defines a local vector in the data space along which the
AP varies monotonically. (Only once such a separation has been achieved can
this vector be calibrated in terms of the AP.)  When
determining multiple parameters (e.g.\ \teff\ and extinction), it is
furthermore essential that the local vectors for each parameter are
near orthogonal, otherwise a local AP degeneracy exists.  Thus
`separation' of sources in HFD must be understood in this more
general, vectorial sense.  Section~\ref{fitness} describes how a
figure-of-merit is constructed to respect these requirements.

HFD is used in section \ref{application} to design filter systems for
the Gaia Galactic Survey Mission and their performance is compared to
other proposed systems. Gaia is a high precision astrometric and
photometric mission of the European Space Agency to be launched in
2010. Operating on the principles of Hipparcos, but exceeding its
capabilities by orders of magnitudes, Gaia will determine positions,
proper motions and parallaxes for the $10^9$ stars in the sky brighter
than V=20 (ESA \cite{esa00}; Perryman et al.\ \cite{perryman01}).  Its
primary objective is to study the structure, formation and evolution
of our Galaxy. To achieve this, the kinematical information must be
complemented with multi-band photometry to determine physical stellar
parameters.  HFD is used to design appropriate UV/optical/NIR (i.e.\
CCD) photometric systems for this survey.  Section \ref{discussion}
then gives a critical discussion of the HFD approach, its features and
limitations and discusses how the approach could be extended and
approved.  Section \ref{conclusions} summarises the main results and
conclusions of this work.

\section{Evolutionary algorithms}\label{EAs}

 Many optimization problems can be viewed as the task
of finding the values of the parameters of a data model which maximize
(or at least achieve a sufficiently large value of) some objective
function.  Deterministic gradient-based methods are
often used, but a major drawback is that they may only sample the
parameter space local to the starting point and thus may only find a
(suboptimal) local maximum.  To overcome this, stochastic optimization
methods can be employed in which random (but not arbitrary) steps are
taken.

One such method draws upon ideas of natural selection found in
biological evolution. In these methods -- collectively known as {\it
evolutionary algorithms} -- a population of individuals (candidate
solutions) is evolved over many generations (iterations) making use of
specific {\it genetic operators} to modify the genes (parameters) of
the individuals. The goal is to locate or converge on the maximum of
some {\it fitness} function (figure-of-merit).  As a population-based
method, it takes advantage of evolutionary behaviour (breeding, natural
selection, maintenance of diversity etc.) to perform more efficient
searches than single solution methods (e.g.\ simulated annealing).

A fairly generic evolutionary algorithm (EA) proceeds as follows.  We
start with an initial population of $\mu$ individuals, perhaps
generated at random. From these, we generate an intermediate
population of $\lambda$ individuals ($\lambda \geq \mu$) either via
{\it recombination} -- the breeding of individuals to produce
offspring with different combinations of their parameters -- and/or
via {\it mutation} -- the application of small random changes to
individuals' parameters.  The $\mu$ fitter individuals from this
intermediate population are then selected.  This {\it selection} could
be carried out deterministically -- take the best $\mu$ -- or
probabilistically, e.g.\ by selecting (with replacement) individuals
from the parent population with a probability proportional to their
fitness. The procedure is then iterated. At any generation we have
$\mu$ different solutions to our optimization problem. Just as is
believed to take place in biological evolution, the evolution of the
system is not directed step-by-step, but rather the population as a
whole improves itself through the constant reproduction of new
individuals and the natural selection of the fitter ones.  A brief
discussion of the different types of EAs plus references to the literature 
is given in Appendix B. The
specific genetic operators used in HFD are described in the next
section.

\section{The HFD model}\label{hfd_model}

\begin{figure}
\centerline{
\psfig{figure=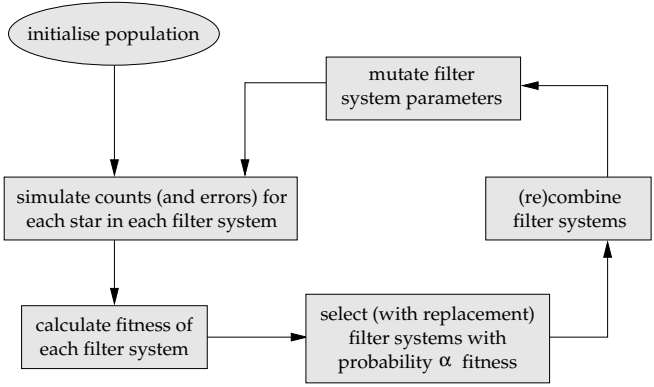,angle=0,width=0.48\textwidth}
}
\caption{Flow chart of the core aspects of the HFD optimization
algorithm. A single loop represents a single iteration, i.e.\ the
production of one new generation of filter systems.}
\label{hfd_flowchart}
\end{figure}

The goal of HFD is to design a survey photometric system according to
how well it can separate stars with different astrophysical parameters
(APs) and avoid degeneracy between the APs.  The filter system is
developed for a specific survey, the scientific goals of which are
represented by a grid of stars with specified APs, magnitudes and
spectral energy distributions (SEDs).  From these and the instrument
model, HFD calculates the fitness of each filter system and uses this
to evolve the population.  This iterative optimization procedure is
summarized in Fig.~\ref{hfd_flowchart}.  The critical aspects of HFD
are now described: the filter system representation
(section~\ref{representation}); the fitness measure
(section~\ref{fitness}); the genetic operators
(section~\ref{operators}).

\subsection{Filter system representation and instrument assumptions}\label{representation}

To be amenable to optimization, the filter system must be
parametrized, or `represented'.  This representation is influenced by
constraints, or fixed parameters, within which the
optimization proceeds. These relate primarily to the instrument.

I first assume that the aperture size (primary mirror area) is fixed
according to financial, technical and other scientific constraints.
Second, I assume that (a) a fixed total amount of integration time is
available for each source to be observed, and that (b) this is the
same for each source. The total integration time per source depends
upon the survey duration, the field-of-view of the instrument, the
area to be observed, and the scanning law (how the field-of-view maps
the sky with time). A uniform scanning law ensures conditions (a) and
(b). Although this is not actually true for Gaia, it is a reasonable
simplifying approximation.  Third, I assume that the total integration
time per source must be divided among all filters. This is the case
for Gaia (and, for example, SDSS) in which the focal place is covered
with a two dimensional array of CCD detectors arranged in one
dimensional strips, with different filters attached directly to
different strips. As the instrument scans the sky the stars cross the
focal plane perpendicular to these strips and the CCDs are clocked at
the same rate.  The integration time in each filter is therefore set
by the width of its respective CCD strip.  Finally, the wavelength
response of the instrument (throughput) and the detectors (quantum
efficiency) are specified and held constant during the optimization.

Each filter in a filter system is parametrized by the following
three parameters: the central wavelength, \cw, the half-width at half
maximum (HWHM), \hwidth, and the fractional integration time, \itime,
i.e.\ the fraction of the total integration time (per source)
allocated to this filter.  The profile of every filter is given by the
generalized Gaussian
\begin{equation}
\Psi(\lambda) = \Psi_0 \,\, {\rm exp}\left[ -(\ln2) \left|\frac{\lambda - c}{b}\right|^{\gamma} \right] \ \ \ .
\label{profile}
\end{equation}
This is Gaussian for $\gamma=2$, and rectangular for
$\gamma=\infty$. Values between these give Gaussian-like profiles but
with flatter tops and steeper sides.  $\gamma=8$ is adopted
throughout.  A fixed peak throughput of $\Psi_0 = 0.9$ is used.

For a system of $I$ filters, there is a total of $3I$ filter system
parameters: these are the free parameters with respect to which the
optimization is performed.  More complex filter parametrizations are
of course possible. For example, additional parameters could allow the
shape, steepness or asymmetry of each profile to be optimized. The
philosophy adopted here is to use a simple parametrization consistent
with a reasonably realistic profile.

The fractional integration time is taken as a real number, $0.0
\leq$~\itime$_i$~$\leq 1.0$, and must of course be normalized,
$\sum_i$\itime$_i = 1.0$. Realistically, CCDs would not be used with
arbitrary widths and hence arbitrary values of
\itime. This could be accommodated by rounding final values of \itime\
or by using a discrete representation (see
section~\ref{nominal_bbp}). 

Given the filter profiles and the fixed instrument parameters, the
number of photons detected in each filter from each source are
simulated. The fitness function also requires the expected photometric
noise.  HFD assumes three noise sources: (1) Poisson noise from the
source; (2) Poisson noise from the background (two contributions: one
over the source and the other arising from the need to do background
subtraction); (3) CCD readout noise.

In the present implementation, the number of filters, $I$, is
fixed. HFD could be generalized to optimize this at the expense of
algorithm complexity. But as we generally consider small
values of $I$ (ca.\ 5--20), I simply run separate optimizations for
different values of $I$ and compare the fitnesses. Furthermore, during
the optimization HFD is able to `turn off' filters by assigning
\itime=0 to a filter, thus reducing the number of effective filters.

\subsection{Fitness measure}\label{fitness}

The fitness measure is a vital part of the optimization procedure and
was qualitatively described in section \ref{intro}.  Not only must it
characterize how well the filter system performs, but it must do this
in such a way that it is appropriately sensitive to all of the
APs.  In fact, the most significant challenge in constructing a
fitness function for this problem was taking into account {\it
multiple} astrophysical parameters, in particular parameters which
have very different magnitude effects on the data and which may be
degenerate. This is certainly the case for the four APs considered in
the Gaia application, \teff, \logg, \feh\ and \extinct.

The fitness function can be considered in three parts, the {\it
SNR-distance}, the {\it AP-gradient} and the {\it orthovariance},
which I now describe.

With $I$ filters, the $R$ sources in the grid define $R$ points in an
$I$-dimensional space, the {\it data space}.  Let the expected (i.e.\
noise free) number of photons detected from the $r^{\rm th}$ source in
the $i^{\rm th}$ filter of the $k^{\rm th}$ filter system be
$p_{k,i,r}$, and let the standard deviation in this (from the noise
model) be $\sigma_{k,i,r}$. These values are normalized to sources of
equal brightness (see below).  The {\it SNR-distance} between source
$r$ and a neighbouring source $n$ is defined as
\begin{equation}
d_{k,r,n} = \sqrt { \left[ \sum_{i=1}^{i=I} 
  \frac{ \left(p_{k,i,r} - p_{k,i,n} \right)^2 }{ \sigma_{k,i,r}^2 + \sigma_{k,i,n}^2 }  
  \right] }   \ \ \ .
\label{SNR-distance}
\end{equation}
Without the denominator, this expression would simply be the Euclidean
distance between sources $r$ and $n$. The denominator
modifies this distance to be in units of the combined noise of the two
sources (standard deviations added in quadrature in the case of small,
uncorrelated noise). 
If we were designing a filter system to discretely classify two or
more classes of objects, 
a suitable fitness measure would be the average
of $d_{k,r,n}$ over all non-similar
neighbours, $n$, (i.e.\ all sources in the other classes), and summed over
all sources, $r$.

The SNR-distance is defined in terms of normalized photon counts to
ensure that it is zero for identical SEDs differing only in apparent
magnitude. 
Source SEDs are normalized such that they all have the same counts in
the G band, the `white light' band used for the astrometric
instrument on Gaia (see section~\ref{Gaia}).

We now need to introduce sensitivity to the APs. For a given
SNR-distance between two stars, $r$ and $n$, the larger their AP
difference, the less fit is the filter system. 
This is quantified with the {\it
AP-gradient}, which for the $j^{\rm th}$ AP is defined as
\begin{equation}
h_{k,r,n,j} = \frac{ d_{k,r,n} }{ | \Delta \phi_{j,r,n} |}
\label{AP-gradient}
\end{equation}
where $\Delta \phi_{j,r,n} = \phi_{j,n} - \phi_{j,r}$ is difference
in the $j^{\rm th}$ AP between sources $r$ and $n$.  To remove the
absolute units of the APs, each AP is scaled to lie in the range 0--1.

We might think that we could generalize eqn.~\ref{AP-gradient} to
account for multiple APs by simply summing over $j$, suitably
weighting each term to account for the fact that small changes in some
parameters (e.g.\ \teff\ and extinction) produce larger changes in the
SED than others (e.g.\ \feh\ and \logg). However, this does not
address the degeneracy between the APs, i.e.\ it ignores the fact that
changes in $d_{k,r,n}$ introduced by varying one AP can be replicated
by varying another AP. Such a measure would therefore be blind to the
{\it individual} effects of each AP on the SED.  As all four APs
considered here have broad band (pseudo-)continuum effects on the SED
(see Fig.\ \ref{sedplot}) this is a significant issue.\footnote{In
principle, narrow band filters measuring individual lines sensitive to
specific APs could overcome some degeneracies, but this is unlikely to
be acceptable for a large, deep survey. Moreover, a fitness measure
explicitly sensitive to AP-degeneracy can quantify this, as
demonstrated in section~\ref{results_mbp}.}

\begin{figure}
\centerline{
\psfig{figure=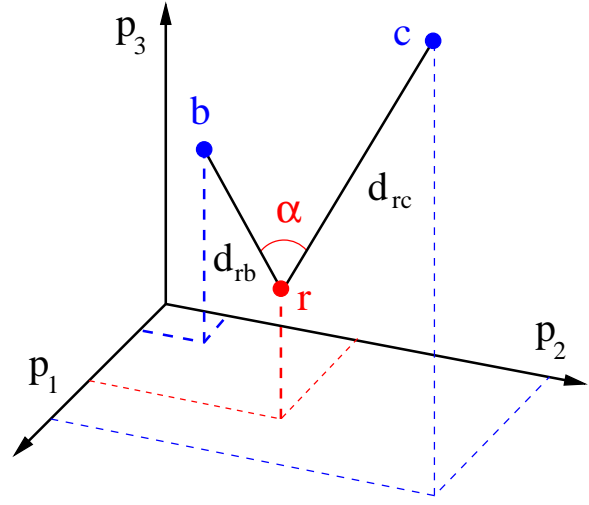,angle=0,width=0.43\textwidth}
}
\caption{Principle of orthovariance
illustrated for a three dimensional data space $p_1$,$p_2$,$p_3$.
For a given source, $r$,
the neighbours $b$ and $c$, which differ from $r$ only in
astrophysical parameters (APs) $j$ and $j'$ respectively, are found.
These are {\it isovars} (``isolated variance'') of $r$ for
APs $j$ and $j'$ respectively.  The vectors $\vec{rb}$ and $\vec{rc}$
are local linear approximations to the {\it principal directions},
those directions in which the APs vary at $r$.  The
angle between the vectors is $\alpha$. The closer to orthogonal these
vectors are the better the vector separation (and hence the lower
the degeneracy) between these two APs at $r$.}
\label{hfd_dist2}
\end{figure}

A filter system free of this degeneracy is one in
which the direction in the data space in which one AP varies (locally)
is orthogonal to the directions in which all other APs vary.  I call
these local vectors the {\it principal directions}, demonstrated for a
three dimensional data space in Fig.\ \ref{hfd_dist2}.
These directions are approximated by the vectors $\vec{rb}$ and
$\vec{rc}$ connecting source $r$ to neighbouring sources $b$ and $c$
respectively. Source $b$ differs from $r$ only in AP $j$; source $c$
differs from $r$ only in AP $j'$.  (A source which
differs from $r$ in only one AP is called an {\it isovar} -- for
``isolated variance'' -- of $r$.)  Let the angle
between these two vectors be $\alpha_{r,j,j'}$. The nearer this angle
is to 90\deg, the lower the degeneracy between APs $j$ and $j'$ at
$r$, and thus the better the filter system.  The nearer
$\alpha_{r,j,j'}$ is to 0\deg\ or 180\deg, the poorer the filter
system is at distinguishing between the effects of the APs, no matter
how large the AP-gradients. Hence a suitable fitness measure could be
proportional to $\sin\alpha_{r,j,j'}$.

This concept, which I call {\it orthovariance}, can be
extended to any number of APs. For $J$ APs we have $J(J-1)/2$ unique
pairings of principal directions at a point and this number of
$\sin\alpha$ (orthovariance) terms.

We now have two distinct figures-of-merit for the performance of a
filter system: the AP-gradients and the orthovariance terms.  For the
single objective optimization approach of HFD, these need to be
combined into a single fitness function. This is done as follows.
The fitness of filter
system $k$ on source $r$ is defined as
\begin{equation}
f_{k,r} = \sum_{j,j' \neq j} x_{k,r,j,j'} 
\label{sourcefitness}
\end{equation}
which consists of $J(J-1)/2$ terms given by
\begin{eqnarray}
\label{crossproduct}
x_{k,r,j,j'} &=& h_{k,r,n_j} \, h_{k,r,n_{j'}} \, {\sin \alpha_{k,r,j,j'}} \\
&=& \frac{ d_{k,r,n_j} \, d_{k,r,n_{j'}} \, {\sin \alpha_{k,r,j,j'}} }{ |\Delta \phi_{j,r,n_j}| \, |\Delta \phi_{j',r,n_{j'}}|} \ \ \ .
\label{vecsep}
\end{eqnarray}
The term $d_{k,r,n_j}$ is the SNR-distance, where $n_j$ means the
nearest neighbour to $r$ which differs only in AP $j$ (i.e.\ source $b
\equiv n_j$ in Fig.\ \ref{hfd_dist2}), and similarly for
$d_{k,r,n_{j'}}$. `Nearest' is in terms of the SNR-distance.  The form
of eqn.~\ref{vecsep} is motivated by the observation that the
numerator is simply the magnitude of the cross product between the two
vectors $\vec{rb}$ and $\vec{rc}$, with the denominator converting
these vectors to AP-gradients.  The sum in eqn.~\ref{sourcefitness} is
over all pairs of APs. As the AP gradients are calculated using
neighbours which differ in only one AP, they are simply the first
order difference approximations of the derivatives of the SNR-distance
with respect to each AP at point $r$ in the grid.  The nominal fitness
for filter system $k$ is then the sum over all sources
\begin{eqnarray}
F_k &=& \sum_r f_{k,r} \\
    &=& \sum_{j,j' \neq j} \sum_r x_{k,r,j,j'} \ \ \ .
\label{nominalfitness}
\end{eqnarray}

As the sources are synthetic spectra, they can be set up on a
sufficiently regular grid to ensure that most sources have isovars.
Some sources may not have isovars for some APs, in which case those
terms in eqns~\ref{sourcefitness} and~\ref{nominalfitness} cannot be
calculated and are omitted.  This is the case for some sources/APs in
the grid used later (Table~\ref{grid}).

Eqn.~\ref{nominalfitness} could be used directly as the fitness
function if it were not for the fact that it suffers from two
problems: AP dominance and a low sensitivity to orthogonality.

To address the former we must appreciate that some APs have a more
pronounced effect on the data than others, i.e.\ a given $\Delta\phi$
for some APs (\extinct\ and \teff) will produce a much larger change
in the SNR-distance than other APs (\logg\ and \feh).  (Recall that
$\phi$ is scaled to the range 0--1 for each AP.) Thus $f_{k,r}$ and
hence $F_k$ will be dominated by a subset of APs and will show little
sensitivity to others, with the result that filter systems are
optimized essentially in ignorance of these `weaker' APs. This may be
overcome by multiplying each AP-gradient term by a factor, $w_j$, to
bring the AP-gradients for each AP to a common level. These factors
are determined by examining the distribution of the AP-gradients for
typical filter systems produced by HFD. Even with these, the fitness
may be dominated by large values of the AP-gradient for a few sources
(so-called `overseparation'). To mitigate this, the AP-gradients may
be raised to a power $1/n$ for $n>1$; $n=2$ is used.

The second problem which arises is as follows. While the cross product
interpretation of eqn. \ref{vecsep} is appealing, it overlooks the
fact that, for example, values of $\sin \alpha$ up to 0.95 occur for
angles only up to 72\deg, yet, intuitively, vectors separated by
90\deg\ ($\sin \alpha = 1.0$) should be considerably more than
1.0/0.95 times fitter. Consequently, I down weight values of
$\sin\alpha$ less than 0.95.
This is done with a two-component
linear transfer function, consisting of a line
joining $(0,0)$ to $(x_0,y_0)$ and another joining $(x_0,y_0)$ to
$(1,1)$, i.e.\
\begin{equation}
\begin{array}{lll}
T(\sin \alpha) & = \ \left(\frac{y_0}{x_0}\right) \sin \alpha & \ \ \  \rm{if} \ \ \ \sin \alpha<x_0 \\
  & = \ \left(\frac{1-y_0}{1-x_0}\right) (\sin \alpha - x_0) \, + \, y_0 & \ \ \ \rm{otherwise}
\end{array}
\label{transfunc1}
\end{equation}
with the transition point $(x_0,y_0) = (0.95,0.1)$.   Generally
speaking, the value of the transmission point, $x_0$, should depend on
both the dimensionality of the data space, $I$, and the number of
principal directions, $J$, because the occurrence and extent of
degeneracies depends on these. For simplicity this fixed
value is used throughout this article. 

Incorporating these two modifications and converting the sums to be
averages (to make the fitness invariant with respect to the number of
sources),\footnote{As some sources do not have isovars they are
dropped from the summation over $r$ and the normalization factor, $R$,
is correspondingly reduced.} the final fitness measure is
\begin{equation}
F^{\ast}_k \ = \ \frac{2}{J(J-1)} \ \sum_{j,j' \neq j} \ \frac{1}{R} \ \sum_r 
x^{\ast}_{k,r,j,j'}
\label{fitmeas}
\end{equation}
{\rm where}
\begin{displaymath}
x^{\ast}_{k,r,j,j'} \ = \ 
 w_j  (h_{k,r,n_j})^{1/2} 
\, w_{j'} (h_{k,r,n_{j'}})^{1/2}
\, T(\sin \alpha_{k,r,j,j'})
\end{displaymath}
where the scale factors are normalized such that $\sum w_j = J$.  The
fitness has units of an AP-gradient per source per AP pair multiplied
by a dimensionless orthovariance factor.

Some aspects of this modified fitness function may seem ad hoc.
However, it was found through detailed experimentation that such
modifications were necessary to increase the sensitivity of the
fitness function.  Further discussion of this point is given in
Appendix A.

Some properties of the fitness should be noted. In the limit
where Poisson noise from the source is
dominant (the `bright star limit'), the SNR-distance, $d$, is scale
invariant with respect to the number of filters, $I$, in the sense
that for a flat spectrum and CCD/instrument response
the fitness is independent of the number of (equal HWHM) filters.
This is relevant because it means that the fitnesses of filter systems
with different numbers of filters can be directly compared. 
In the bright star limit, the SNR-distance is linearly proportional to
the SNR. As long as this limit holds, the HFD optimization is
independent of source magnitude (because the genetic selection
operator is invariant with respect to multiplicative scalings of
the fitness).
In the faint star limit -- where source-independent noise terms
dominate -- $d \propto 1/\sqrt{I}$. In other words, at faint
magnitudes there is a penalty to be paid for retaining a large number
of filters, something which is seen in the applications in
section~\ref{application}.


\subsection{Genetic operators}\label{operators}

The evolutionary aspects of EAs which distinguish them from random
searches are embodied in the genetic operators.  The principal
operators are {\it selection}, {\it recombination} and {\it mutation}
(see Fig.~\ref{hfd_flowchart}). Between them, these operators provide
for an exploration of the parameter space (mutation and recombination)
and an exploitation of the fitter solutions (selection).

\subsubsection{Selection}\label{selection}

Selection is performed probabilistically via the commonly-used
`roulette wheel' method: Each individual is selected from the parent
population with a probability proportional to its fitness
(eqn.~\ref{fitmeas}).
As selection is done with replacement, the expectation is that
individuals are selected with a frequency proportional to their
fitnesses.  Note that we do not simply retain the fittest individuals
at each generation. While this would guarantee a monotonic increase in
the maximum fitness, it would rapidly erase diversity in the
population resulting in premature convergence to a poor local maximum.
Even so, with a finite population there is a chance that the best
individuals are not selected and that improvements in fitness from
earlier generations are lost. To guard against this, {\it elitism} is
used: the $E$ fittest individuals ($E < K$, where $K$ is the
population size) are copied to the next generation without
modification.  The remaining $K-E$ individuals for the next generation
are selected probabilistically from the full parent population
(including the $E$ elite), and combined/mutated in the normal way. We
shall see in section~\ref{variations_bbp} that elitism produces
significant performance gains.


\subsubsection{Recombination}\label{recombination}

After an individual has been selected, it is (re)combined with a
probability \Precomb\ with a second selected individual.  This is done
by randomly selecting one filter from each system and swapping them.
A value of \Precomb=1/3 is used on the basis that the
expected fraction of offspring produced by recombination is then 0.5.
It turns out (section~\ref{variations_bbp}) that HFD is very
insensitive to \Precomb\ and that this operator is actually
unnecessary.

\subsubsection{Mutation}\label{mutation}

After selection (and possibly recombination) each parameter (\cw,
\hwidth\ and \itime) of each filter is mutated with
a probability \Pmut.  A mutation is a random Gaussian perturbation.
For the central wavelength, the mutation is additive: \cw\
$\rightarrow$ \cw~$+$~N(0,\,\sigcw).  For the filter width and
fractional integration time, the mutation is multiplicative: \hwidth\
$\rightarrow$ \hwidth($1.0~+$~N(0,\,\sighwidth)) and \itime\
$\rightarrow$ \itime($1.0~+$~N(0,\,\sigitime)). As \sighwidth$<1.0$,
\sighwidth\ can be considered as the typical fractional change in
\hwidth, and likewise for \itime. Whereas linear changes in the
central wavelength seem an appropriate way of sampling that parameter,
changes proportional to the current size of the parameter seem more
appropriate for the HWHM and fractional integration time.

HFD operates within both absolute wavelength limits defined by the
CCD/instrument profile and wavelength limits on the HWHM, such that
mutations which would violate these limits are not accepted (see
section~\ref{modelloops}). Limits are also applied to the fractional
integration time. A minimum is applied such that if a mutation sets a
value of \itime\ below \itimemin\ then \itime\ is set to zero.
The filter can be turned on again by any successful positive
mutation.  Thus while the number of filters in the model is fixed, the
number of {\it effective filters}, i.e.\ filters with \itime$>$0, is
variable.  This lower limit on \itime\  was imposed to prevent very short
integration times, which would require unrealistically narrow CCDs.
Likewise, mutations which would take \itime\ above \itimemax\ are
rejected.  For $I$ filters, I use \itimemin$=1/(4I)$ and
\itimemax$=4/I$ but with the latter truncated to a maximum of 0.5.
Experimentation has shown that this upper limit on \itime\ is probably
not necessary, because the fitness function itself penalizes such
solutions through the lack of integration time it permits for other
filters.

For the evolutionary search mechanisms to be superior to a random
search we must assume that the fitness is a smooth function of the
filter system parameters over some reasonable length scale.  The
mutation sizes should be comparable to these length scales; if they
were much larger then the childs' fitness would not correlate with its
parents' fitness and the search would be quasi-random.  Quantifying
these length scales is not straight forward without knowledge of the
shape of the fitness landscape. A typical mutation size should
obviously be much smaller than the total range of a parameter and from
our astrophysical knowledge we can also say that mutations below some
value will have negligible effect. Based on such considerations as
well as experimentation, the values of \sigcw, \sighwidth\ and
\sigitime\ were fixed at 500\,\AA, 0.50 and 0.25
respectively. Experimentation has found that the results of HFD are
not very sensitive to these values (see section~\ref{variations_bbp}).
Likewise, the evolution is not very sensitive to \Pmut, which was set
to 0.4.

\subsection{HFD initialization and execution}\label{modelloops}

The population is initialized by drawing \cw\ and \hwidth\ at random
from a uniform distribution between the minimum and maximum values of
these parameters.  The permissible wavelength range is determined by the
CCD/instrument response (Fig.~\ref{responses}), and is is set to
2750--11\,250\,\AA\ for BBP and 1750--11\,250\,\AA\ for MBP (the two
instruments on Gaia; see section~\ref{Gaia}). These are extended
compared to the zero response values to permit cut-off filters.  The
permitted range for the HWHM is set at 80--4000\,\AA. The lower limit
is introduced to avoid errors interpolating the SEDs (and very narrow
filters are anyway not acceptable on SNR grounds). The upper limit is
essentially no limit as it encompasses the entire permissible
wavelength range.  Interestingly, the optimization naturally constrains
itself to a more limited range of HWHM (section~\ref{application}).
The fractional integration times are initialized to be equal (to
$1/I$).

The evolution is terminated after a fixed number of generations,
typically 200, beyond which the rate of increase of fitness in
numerous configurations was found to be very small.  This entire
optimization process is repeated for a number of {\it runs} commencing
from different initializations to investigate how consistently HFD
converges on a common solution (sensitivity to initial conditions).

The parameters involved in HFD are summarized in
Table~\ref{param_table}, where the values for the `nominal'
optimizations in section~\ref{application} are also given.  What
little theory exists to guide us in setting the parameters of the EA
is derived from very simple problems which may have little
generality. One is therefore forced to perform tests and build up
experience of the sensitivity of the model to these parameters.  A
population size of 200 and an elite of 10 was selected somewhat
arbitrarily: the effect of varying these and other parameters will be
discussed below.


\begin{table*}
\begin{minipage}{0.85\textwidth}
\caption{HFD parameter overview. The optimization is done with respect
to the $3I$ free parameters (with bounds). Symbols for parameters are
given where used in the text. The nominal parameter values used in the
simulations described in section~\ref{application} are listed. The two
sets of instrument parameters labelled ``BBP'' and ``MBP'' refer to
the two Gaia instruments described in section~\ref{Gaia}.  }
\label{param_table}
\begin{tabular}{l}
\\
{\bf Free parameters}\\
$
\left. \begin{array}{p{8cm}l}
\textrm{central wavelength / \AA}           & \textrm\cw \\
\textrm{full-width at half maximum / \AA}   & \textrm\hwidth \\
\textrm{fractional integration time}        & \textrm\itime \\
\end{array} \right\}\times I
$ 
\end{tabular} \\
\begin{tabular}{p{8cm}p{1cm}p{2cm}}
~\\
\multicolumn{2}{l}{{\bf Fixed parameters: fitness measure}} \\
~~stellar population (number of sources, $R$ = 415)      & & Table~\ref{grid_table} \\
~~magnitude of stars (in G band)$^1$                   & & 15 \\
~~AP weight \extinct\         &   & 1.5/128 \\
~~AP weight \feh\             &   & 75.0/128 \\
~~AP weight \logg\            &   & 50.0/128 \\
~~AP weight \logteff\         &   & 1.5/128 \\
~\\
\multicolumn{2}{l}{{\bf Fixed parameters: evolutionary algorithm}} \\
~~number of filter systems ($=$ population size)    & $K$        &  200 \\
~~size of elite                                     & $E$        &  10 \\
~~number of generations                             &            &  200 \\
~~number of runs$^2$                                &            &  20 \\
~~probability of recombination                      & \Precomb   &  1/3 \\
~~probability of mutation                           & \Pmut      &  0.4 \\
~~std.\ dev.\ of mutation for \cw\  / \AA\           & \sigcw\    &  500 \\
~~std.\ dev.\ of mutation for \hwidth\               & \sighwidth\ &  0.5 \\
~~std.\ dev.\ of mutation for \itime\                & \sigitime\ &  0.25 \\
\end{tabular}
\begin{tabular}{p{8cm}p{1cm}p{2cm}p{2cm}}
&\\
\multicolumn{2}{l}{{\bf Fixed parameters: instrumental}} & {\bf BBP} & {\bf MBP}\\
~~filter profile                      &      &  eqn.~\ref{profile} & eqn.~\ref{profile}  \\
~~number of filters                   & $I$  &  5       &  10     \\
~~telescope aperture area / m$^2$     &      &  0.7     &  0.25   \\
~~total integration time / s          &      &  1205    & 16500   \\
~~CCD \& instrument response          &      &  Fig.~\ref{responses} & Fig.~\ref{responses} \\
~~CCD readout noise / e$^{-}$         &      &  251     & 277     \\
~~effective background / G mag        &      &  22.37   & 18.29   \\
~~min.(\cw -\hwidth), max.(\cw +\hwidth) / \AA\      &      &  2750, 11250 & 1750, 11250 \\
~~min.\ \hwidth, max.\ \hwidth\ / \AA\ &     &  80, 4000 & 80, 4000 \\
~~min.\ \itime, max.\ \itime\          & \hspace*{-1.5em}\itimemin, \itimemax\      &  0.05, 0.5 & 0.025, 0.4 \\

\\
\hline 
\multicolumn{4}{p{15cm}}{$^1$~The G band is defined in section~\ref{Gaia}. The stars do not in general have to have the same magnitude.} \\
\multicolumn{4}{p{15cm}}{$^2$~The number of independent runs of HFD from different initializations is not a parameter of the model.} \\
\end{tabular} \\
\end{minipage}
\end{table*}

\section{Application of HFD to the design of Gaia filter systems}\label{application}

HFD is applied to design photometric systems for two Gaia
instruments.  In both cases the goal is to achieve systems which can
best determine the four astrophysical parameters, effective temperature
(\teff), surface gravity (\logg), metallicity (\feh) and interstellar
extinction\footnote{Extinction is not an intrinsic stellar property,
but as extinction can vary considerably on small spatial scales it
should ideally be determined for each star individually.}(\extinct),
for a grid of stars.

\subsection{The stellar grid}\label{grid}

The main purpose of the grid is to sample the dependence of the SED on
the APs in order to calculate the fitness, and from this
perspective it is not necessary to have a very dense grid (but see
section~\ref{discussion}). The grid used is shown in
Table~\ref{grid_table}. It has been constructed loosely considering
the scientific goals of Gaia.  The SEDs are Basel2.2 synthetic spectra
(Lejeune et al.\ \cite{lejeune97}) which were artificially reddened
using the curves of Fitzpatrick (\cite{fitzpatrick99}) with R$_{\rm
V}$=3.1.

For a given optimization, all stars are presented at the same
magnitude in the G band (see section~\ref{Gaia}).  The nominal
optimization is carried out at G=15, the target magnitude for Gaia
(ESA \cite{esa00}). Section~\ref{variations_bbp} demonstrates the
effect of varying this magnitude. Note that the SEDs themselves are
noise free: the magnitude at which they are presented determines the
noise in the SNR-distance (eqn.~\ref{SNR-distance}).  All magnitudes
are on the AB system (Oke \& Gunn \cite{oke83}).

By necessity, this grid is a simplification of the true diversity of
scientific targets which Gaia will encounter.  Many more sources and
additional astrophysical parameters could be included, and done so at
characteristic magnitude ranges. 

\begin{table*}
\begin{center}
\parbox{10.5cm}{
\caption{The stellar grid in \teff\ and \logg\ (spectral types are
given for guidance).  Each of these 17 AP combinations is reproduced
at the five metallicities and five extinctions shown at the bottom of
the table, giving a total of 425 sources (some metallicities are
missing from the Basel library, so there are actually only 415
sources).}
\label{grid_table}
}
\vspace*{2ex}
\begin{tabular}{p{2.8em}p{2.8em}p{2.8em}p{2.8em}p{2.8em}p{2.8em}p{2.8em}p{2.8em}}
\hline
\logg	& \multicolumn{7}{c}{\teff\ / K (SpT)} \\
\hline 
~\\
4.5	& 3500	& 4750	& 5750	&&&& \\			
	& MV	& KV	& GV	&&&& \\		
4.0	&&&			& 6750	& 8500	& 15000	& 35000 \\
	&&&			& FV	& AV	& BV	& OV \\
3.5	&&&&&&& \\
	&&&&&&& \\		
3.0	&&&			& 6000	& 8500 && \\
	&&&			& RRLyr & BHB && \\
2.5	&&		& 5500  &&	        & 15000 & \\
	&&		& GIII  &&		& BIa   & \\
2.0	&	& 4500	& 5500  &&&& \\
	&	& KIII	& FIa   &&&& \\
1.5	&&&& 				& 8500	&& \\
	&&&&				& AI    && \\
1.0	& 3500	&	& 5000	&&&& \\	
	& MIII	&	& GI	&&&& \\	
0.5	&&&&&&& \\
	&&&&&&& \\		
0.0	& 3500 &&&&&& \\
	& MIa  &&&&&& \vspace*{1ex}  \\
\hline \\
& \feh\ & =     & $+$0.5 & 0.0 & $-$0.5 & $-$1.5 & $-$2.5 \\
& \extinct\ & = & 0.0 & 0.2 & 2.0 & 5.0 & 10.0 \\
~\\
\hline
\end{tabular}
\end{center}
\end{table*}

\subsection{The Gaia instrument model}\label{Gaia}

Gaia employs two separate telescopes (focal planes; see
section~\ref{representation}) each equipped with different instruments
(see ESA \cite{esa00} or Perryman et al.\ \cite{perryman01}, although
the designs have since been slightly modified and may well be modified
again).  The first instrument is the astrometric instrument comprising
a large array of unfiltered CCDs. This pass band -- called the {\it G
band} -- is defined by the CCD/instrument response.  The centroid of
the point spread function through this broad band is colour dependent,
so to achieve accurate astrometry a chromatic correction is
required. This is supplied by a number of (typically) broad band
filters on the trailing edge of the focal plane, referred to as the
{\it Broad Band Photometer}, BBP. It turns out that provided there are
four or five filters covering the G
band, we are free to optimize BBP for other purposes (Lindegren
\cite{lindegren01}), e.g.\ stellar parametrization.  As focal plane
area is limited in this instrument, a second instrument, the {\it
Medium Band Photometer}, MBP, exists, the primary goal of which
is stellar parametrization.  It works on the same principle as BBP,
and although it has a smaller telescope aperture than BBP, it has a
much larger area and field of view, so more filters can be
allocated. Current designs for MBP have considered 8--11 bands.

\begin{figure}
\centerline{
\psfig{figure=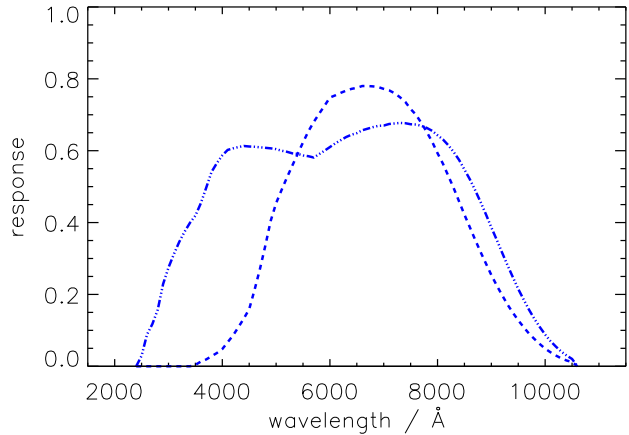,angle=90,width=0.50\textwidth}
}
\caption{Responses (CCD quantum efficiency multiplied by instrument
throughput) for the BBP instrument (dashed line) and the MBP
instrument (dot-dashed line). The instrument responses are six
reflections of silver and three of aluminium respectively.  The BBP
response defines the G band, which, along with the AB system, defines
the magnitude system adopted for Gaia.  The MBP response is a
composite of two different CCD QE curves (joined at 5700\,\AA\ where
they have equal QE).}
\label{responses}
\end{figure}

The instrument models used here for MBP and BBP reflect the Gaia
designs as of mid 2003.  The main parameters are summarized in
Table~\ref{param_table}.  Each instrument has a different wavelength
response and uses different CCDs (Fig.~\ref{responses}). MBP
additionally has red- or blue-enhanced CCDs depending on the filter central
wavelength. For simplicity, a composite of these two is used in HFD by
taking the maximum of each QE profile.
For the noise model, a background with solar SED and V=22.50 (G=22.18)
mag/sqarcsec is assumed (ESA \cite{esa00}). This is translated to
background counts in the source extraction using aperture photometry,
yielding the effective background in Table~\ref{param_table}.  The
large background for MBP is a result of optical aberrations (from the
short focal length) giving rise to poor spatial resolution
necessitating a large extraction aperture. This will also produce
source confusion at brighter magnitudes than occurs with BBP. Due to
these different characteristics, a joint optimization of the MBP and
BBP filter systems is probably not desirable.

While the terms BBP and MBP will be retained for these instruments, it
should be noted that the HFD optimization sets essentially no limits
on the HWHM of the filter profiles (section~\ref{modelloops}).




\subsection{Results: BBP}\label{results_bbp}

The first part of this section examines in detail the optimization of
a 5-filter BBP system using the nominal settings in
Table~\ref{param_table}. The second part examines the effect of
varying these parameters.

\subsubsection{Nominal model parameter settings}\label{nominal_bbp}

\begin{figure}
\centerline{
\psfig{figure=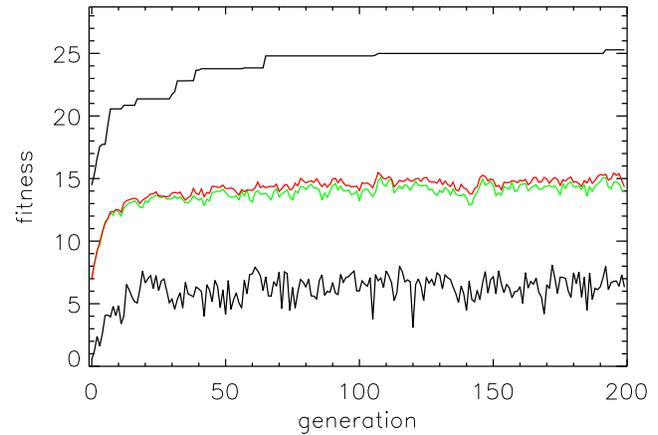,angle=90,width=0.50\textwidth}
}
\caption{Evolution of fitness statistics
for a typical HFD run applied to the nominal BBP model.  The lines
from top to bottom denote the maximum, mean, median and minimum fitness
in the population.}
\label{fit260_5.log}
\end{figure}

The typical evolution of the fitness during an optimization run is
shown in Fig.~\ref{fit260_5.log}.  Starting from the initial random
population, we see a rapid increase in all fitness statistics over the
first few generations followed by a slower increase over the rest of
the evolution. The mean and median, always very close, oscillate
around a constant value after 20--40 generations. The minimum value
shows similar behaviour, but with larger negative dips indicating the
creation of poor solutions. In contrast, the maximum fitness never
decreases, as guaranteed by elitism
(section~\ref{selection}). Significantly, the maximum fitness
continues to increase after the other measures have levelled off,
although by decreasing amounts: While the population as a whole
`stagnates', a few ever fitter individuals continue to be
created. This is what is important, as the goal of the evolutionary
algorithm in this problem is to achieve the highest fitness of a
single individual (the {\it run maximum}), rather than improve the
whole population.
The increase in maximum fitness after 200 generations looks asymptotic.
Extending the evolution to ten times as many generations only
improves the run maximum fitness by around 2\%.

\begin{figure}
\centerline{
\psfig{figure=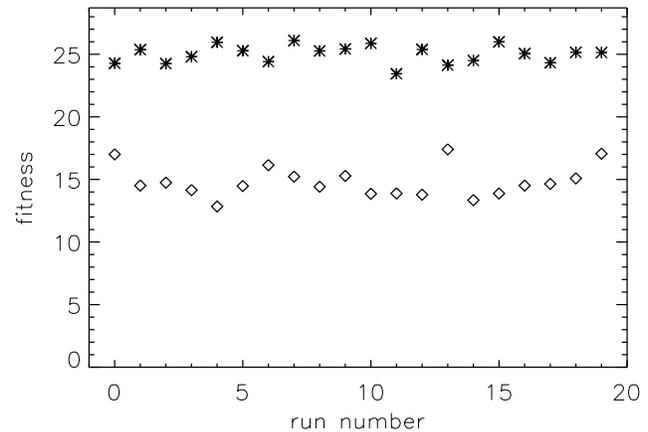,angle=90,width=0.50\textwidth}
}
\caption{Maximum fitness in the initial random population (diamonds) and in
the final population (stars) for each of 20 independent BBP runs.}
\label{fit260_runs.log}
\end{figure}

Fig.~\ref{fit260_runs.log} shows that the run maximum fitness is
fairly consistent across runs, the spread across 20 runs being about
10\%. It further shows that the improvement in maximum fitness as a
result of using the search and selection operators (as opposed to
random search) is a factor of about 1.8.

\begin{figure}
\centerline{
\psfig{figure=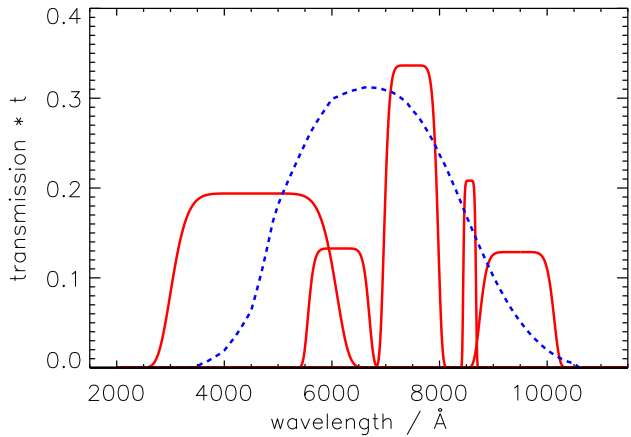,angle=90,width=0.50\textwidth}
}
\caption{The HFD-1B filter system, a BBP filter system produced by HFD
(run no.\ 7 -- the fittest run -- in Fig.~\ref{fit260_runs.log}).  The
peak filter transmissions have been scaled to the fractional integration time
for each filter (the true peak transmission of each filter is fixed at
0.9).  Thus the exact points of overlap of the filters are not
accurately depicted.  The dashed line shows the instrument+CCD
response curve for BBP, arbitrarily scaled in the vertical direction.
The filter parameters are listed in Table~\ref{hfdfs_params}.  }
\label{fs260_7.trans}
\end{figure}

\begin{figure*}
\centerline{
\psfig{figure=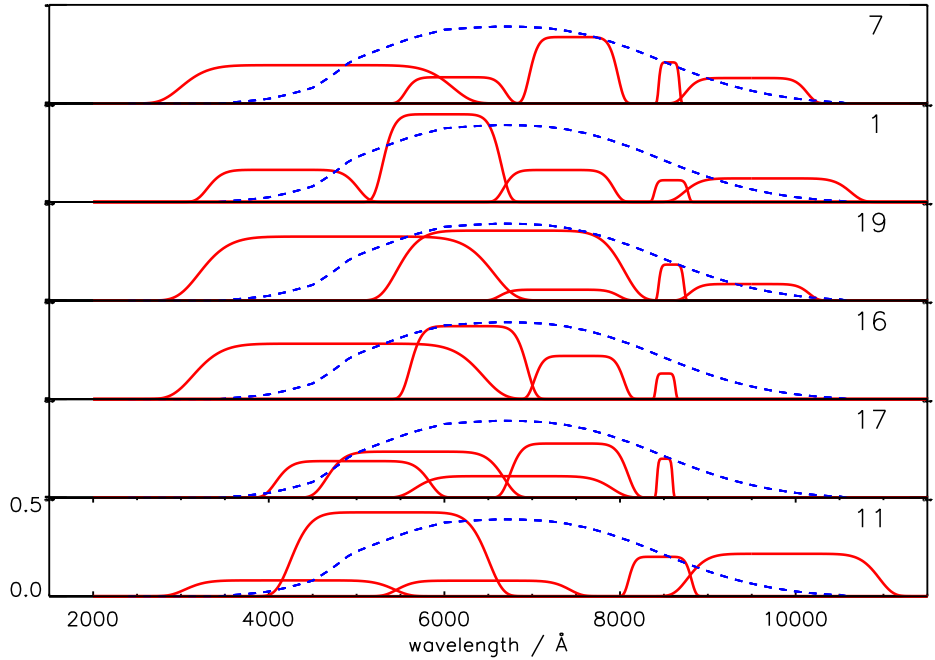,angle=90,width=0.75\textwidth}
}
\caption{The optimized BBP filter systems for six different runs (six
different initializations), selected to show maximum variance between
the filter systems (plotted as in Fig.~\ref{fs260_7.trans}). They are
ordered from fittest (at the top, same as Fig.~\ref{fs260_7.trans}) to
the least fit (at bottom). The run numbers in each panel correspond to
those in Fig.~\ref{fit260_runs.log} }
\label{fs260_runs.trans}
\end{figure*}

Fig.~\ref{fs260_7.trans} shows the best filter system produced from
these 20 runs. A number of features are immediately apparent.  First,
the filters cover the entire wavelength range, and four of
the five filters are rather broad.  Second, the reddest and bluest
filters extend essentially to the longest and shortest wavelengths
possible with the instrument response, i.e.\ they are
cut-off filters. Third, all five filters have non-zero integration
time. These general features are consistently found in the run maxima
of the 20 runs, and the best five runs in particular produce very
similar filter systems.  An indication of the overall consistency is
given in Fig.~\ref{fs260_runs.trans}, where six filter systems
ranging from the fittest to least fit run maxima across the 20 runs
are shown.  More significant differences occur among the less fit
systems, e.g.\ the lack of a red cut-off filter, or more significant
overlap of filters. In two cases there are only four effective filters
(i.e.\ the fifth has \itime=0) and in another case two filters are almost
identical. While HFD appears to converge toward a stable filter
system, no convergence
criterion is used to terminate the evolution. The differences between
the final filter systems (and fitnesses) of each run could, therefore,
reflect incomplete convergence as much as convergence on different
local maxima of the filter parameter space.

A consistent aspect of HFD -- seen for many variations of the grid,
instrument and EA parameters -- is that it produces systems with broad
filters. At first this seems counterintuitive, as we may expect the
best distinction between stellar parameters to be achieved by placing
narrow filters on specific (narrow) features. Inspection of the
fitness function, eqn.~\ref{fitmeas} and in particular the
SNR-distance component (eqn.~\ref{SNR-distance}), gives some
explanation. For constant values of $\sin\alpha$, the fitness can be
increased by making the filters wider. This increases the SNR so is
obviously desirable. If this widening increases the degeneracy between
APs then it is penalized through a reduction in the value of
$\sin\alpha$ and thus a decrease in the fitness.  We can think of HFD
as attempting to simultaneously achieve the largest values of the
SNR-distance (or rather, the AP-gradients) between sources consistent
with also maximizing their vector separation
(Fig.~\ref{hfd_dist2}).  That this is functioning at some level is
indicated by the fact that although the filter half widths can extend
up to (and are initialized up to) 4000\,\AA, in the significant majority of
optimized systems they are less than 2000\,\AA.  Clearly there is an
orthovariance penalty to be paid by much wider filters.

\begin{figure*}
\centerline{
\psfig{figure=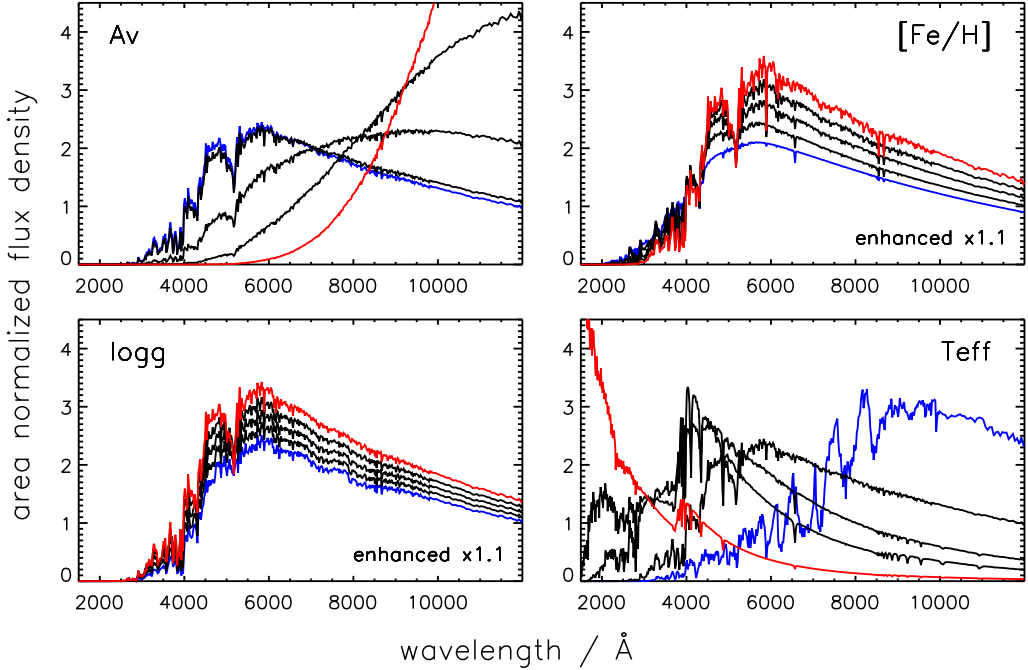,angle=90,width=0.75\textwidth}
}
\caption{Variations of the four APs have broad band effects. Taking
the SED with APs \extinct=0.0\,mag, \feh=0.0\,dex, \logg=4.5\,dex,
\teff=4750\,K, each panel shows the effect of varying one AP over the
full range shown in the grid (Table~\ref{grid_table}).  The magnitude
of the effects of \logg\ and \feh\ have been enhanced for clarity by
multiplying the five SEDs by 1.0,1.1,$\ldots$,1.4.  The SEDs are
plotted to have equal integrated flux density over the
wavelength interval 900--12\,000\,\AA.}
\label{sedplot}
\end{figure*}

Fig.~\ref{sedplot} goes further to explain why broad filters may be
desirable. It shows that the effects on the SED of varying any of the
four APs are coherent over a wide wavelength range and not restricted
to specific, narrow wavelength intervals.  Thus on signal-to-noise
grounds -- and subject to the orthovariance requirement -- broader
filters are more sensitive to AP variations.\footnote{This would not
be true for APs which have a very localized wavelength signature, such as
specific element abundances.}  This can be exploited by Gaia because,
unlike ground-based surveys which are often limited by imperfect
calibration of variable telluric effects, Gaia can make
reliable use of the stellar continuum and unresolved features.

Another property often seen is overlapping filters.  Large amounts of
overlap are not desirable from the point of view of colour-colour
diagrams. However, colour-colour diagrams are probably not the optimal
way of determining stellar parameters. After all, such diagrams only
make use of two or three bands (the normalization already being provided by
the G band), whereas HFD is performing a separation directly in the
higher dimensional space (five dimensions in this case, ten for MBP).  This is
likely to make a more efficient use of multivariate data than do
colour-colour diagrams, which, from the point of view of stellar
parametrization, are only a means to an end.

\begin{figure*}
\centerline{
\psfig{figure=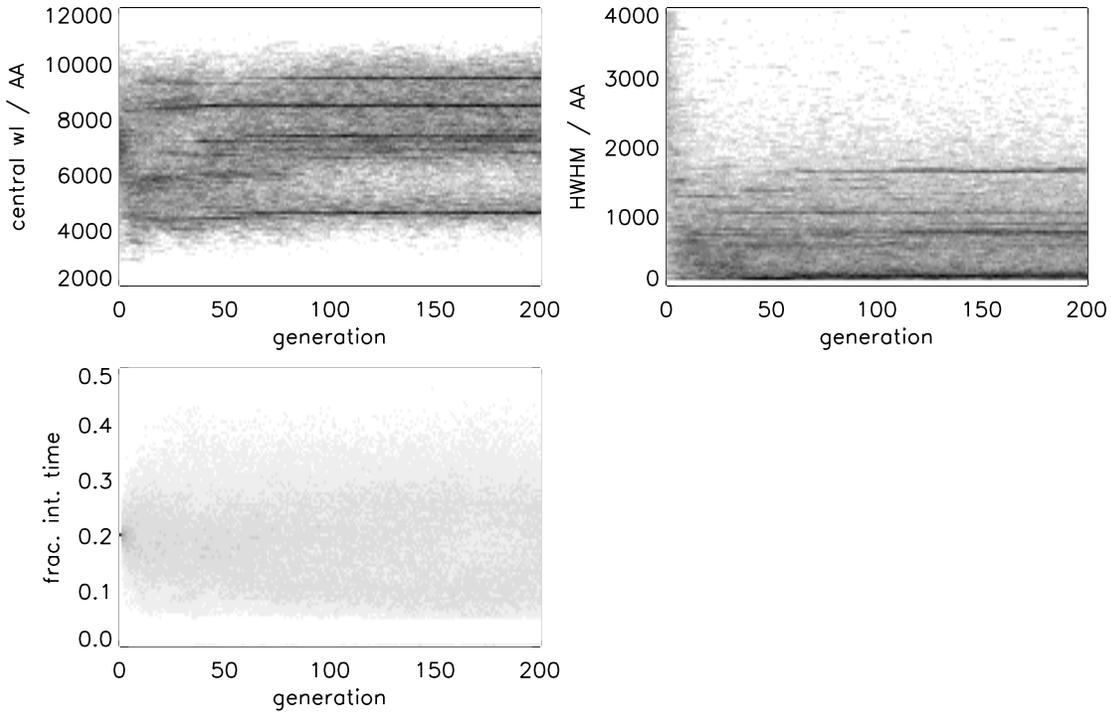,angle=90,width=0.85\textwidth}
}
\caption{Evolution of all filter system parameters for the nominal
BBP setup with $I$\,=\,5 filters for a typical run (the corresponding
fitness evolution is shown in Fig.~\ref{fit260_5.log}).  At
each generation the $I \times K = 1000$ values for that filter
parameter type are plotted as a grey scale (with square root intensity
scaling used to enhance sparse regions).}
\label{fs260_5.log}
\end{figure*}

To get a better idea of how HFD works, we may investigate the
evolution of the filter system parameters. There is immediately a
difficulty here because for $I$\,=\,5 filters there are formally
3$I$\,$-$1\,=\,14 independent parameters, the joint evolution of which
cannot be visualized. Instead, Fig.~\ref{fs260_5.log} shows the
evolution of each parameter type separately for a typical run.  The
central wavelengths occupy the full range of possible values
throughout the evolution, although after 40--70 generations they show
more concentration around a handful of values. Changes can be
correlated with changes in the fitness evolution
(Fig.~\ref{fit260_5.log}).  In contrast to this behaviour, within 10
generations most filters with a HWHM more than about 2000\,\AA\ are
purged from the population only to make short-lived appearances.
 This self-regulation property of HFD to remove very broad filters
is frequently observed. 

The fractional integration time, \itime, shows a much more continuous
distribution between its bounds. This may indicate that the exact
setting of \itime\ is not that critical. To test this I repeated the
set of 20 runs with \itime\ fixed at 0.2 for all filters.
The median fitness of the run maxima is about 4\% lower and
the filter systems are similar in their general properties.
If, instead, the fractional integration times of the fully optimized
system (Fig.~\ref{fs260_7.trans}) are set to 0.2 and the fitness
recalculated, the fitness is found to be about 12\% lower. With BBP on
Gaia, six CCD slots may be available, enabling us to allocate two slots
(i.e.\ \itime=2/6) to the filter with largest \itime\ and 1/6 to the
rest. With this discretization, the fitness decreases by less than 2\%
relative to the full optimization. In conclusion, optimizing \itime\
is desirable, but moderate rounding to match discrete CCD widths may
not significantly degrade performance.

\begin{figure*}
\centerline{
\psfig{figure=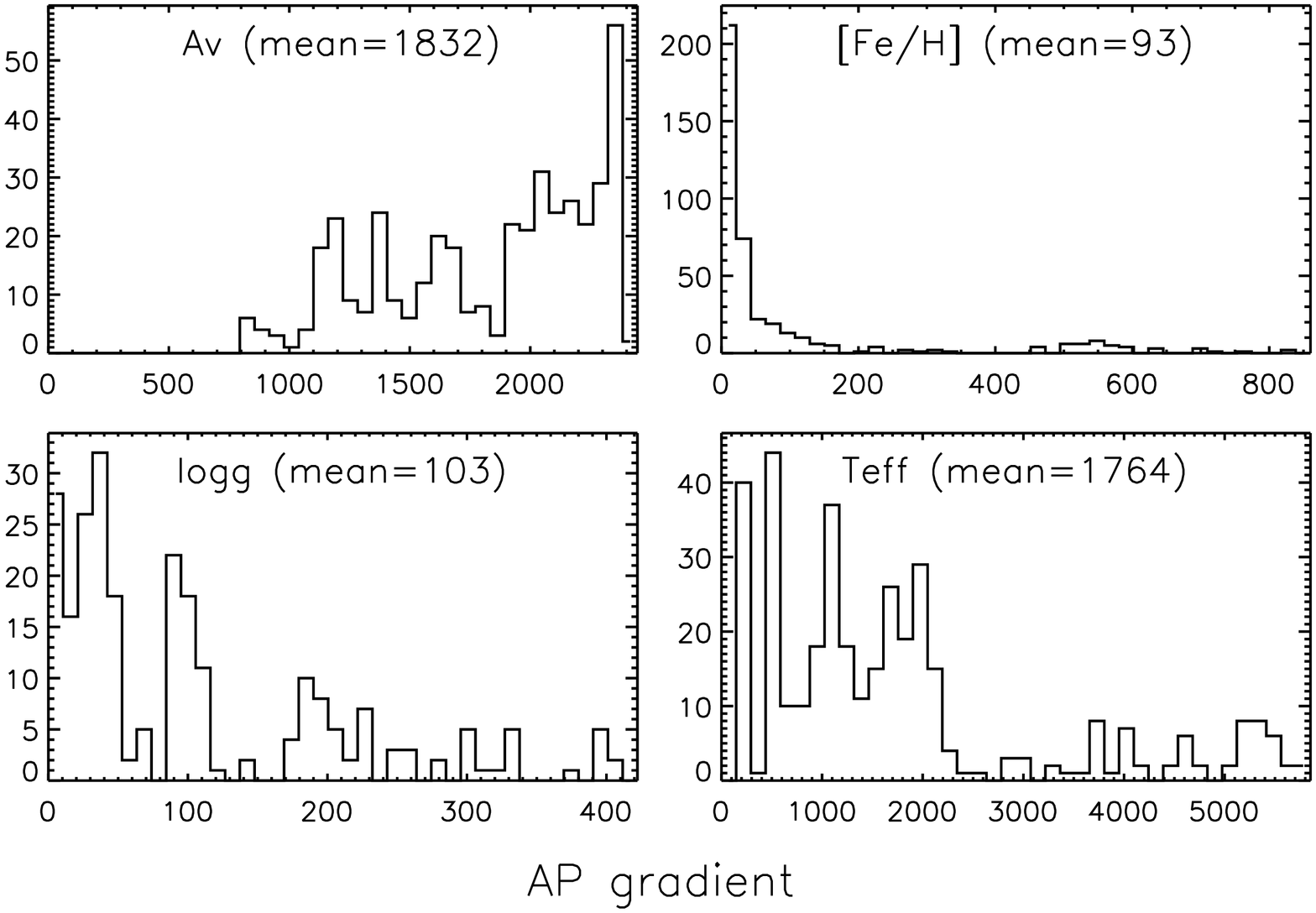,angle=90,width=0.70\textwidth}
}
\caption{Distributions of the four AP-gradients for all sources in the
grid produced by the BBP filter system shown in Fig.~\ref{fs260_7.trans}. The
mean values of the distributions are given.  For comparison, these
values averaged over 1000 random filter systems are: \extinct=1435;
\feh=49; \logg=68; \teff=1461.}
\label{fit260_7.apgrad}
\end{figure*}

\begin{figure*}
\centerline{
\psfig{figure=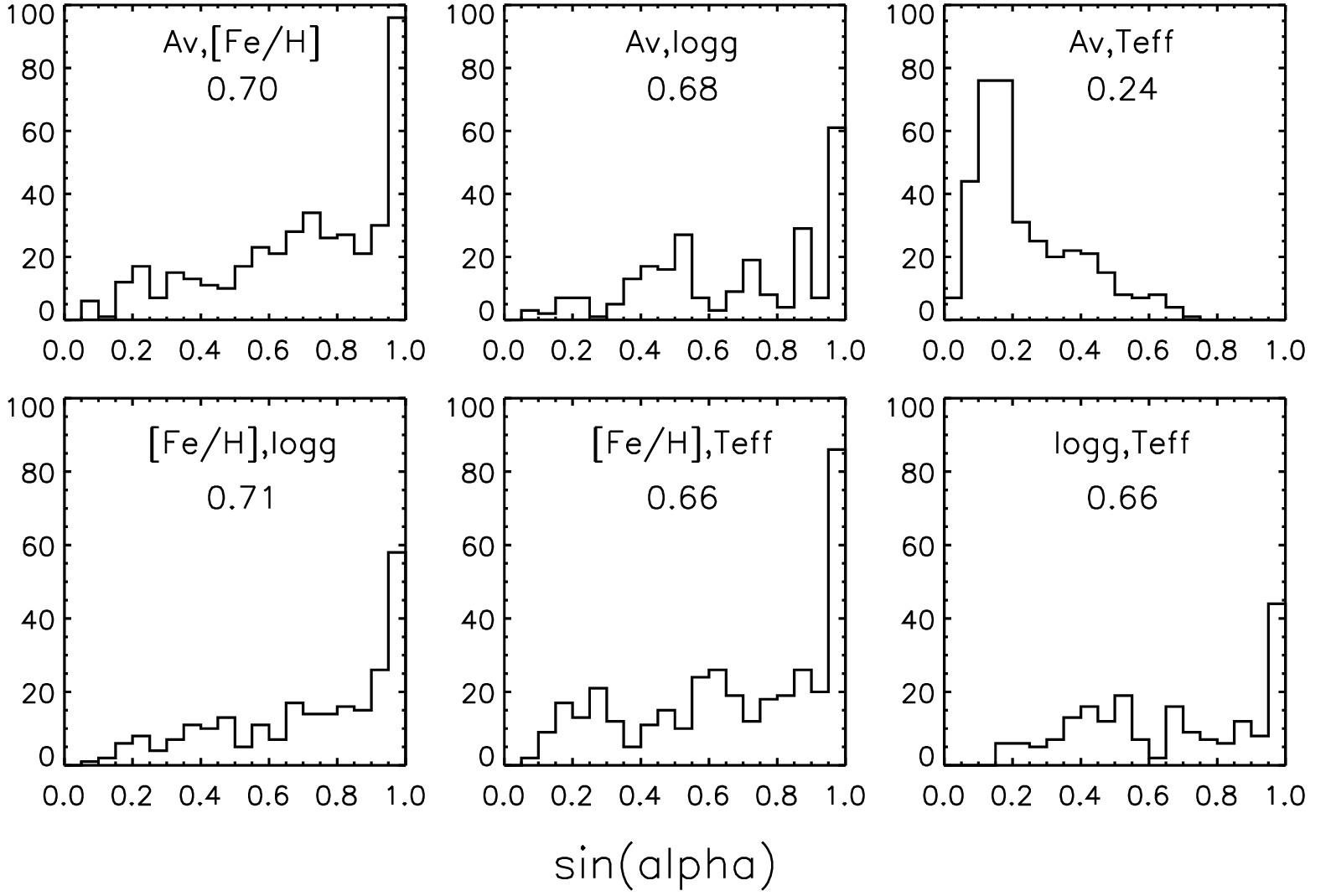,angle=90,width=0.70\textwidth}
}
\caption{Distributions of the six orthovariance terms ($\sin\alpha$)
for all sources in the grid produced by the BBP filter system shown in
Fig.~\ref{fs260_7.trans}. The mean values of the distributions are
given. For comparison, these values averaged over 1000 random filter
systems are: \extinct,\feh=0.52; \extinct,\logg=0.47;
\extinct,\teff=0.18; \feh,\logg=0.51; \feh,\teff=0.50;
\logg,\teff=0.46.}
\label{fit260_7.sinalpha}
\end{figure*}

We can obtain a better idea of the performance of a filter system by
looking at the distributions of the four AP-gradients and six
orthovariance terms comprising the fitness, as shown in
Figs.~\ref{fit260_7.apgrad} and~\ref{fit260_7.sinalpha}. In the
AP-gradient calculation the full range of each AP is normalized to the
range 0--1. Therefore, if we want a 2.5\% difference in APs (e.g.\ 0.1
dex in \feh\ or \logg, or 0.25 dex in \extinct) to be separated by a
SNR-distance of at least 5, then we require AP-gradients of at least
5/0.025=200. We see that this is easily achieved (at this G magnitude)
for \teff\ and \extinct\ for essentially all sources but is not for
\logg\ or, especially, \feh, for many sources. At some level this is
to be expected, since \logg\ and \feh\ are `weak' APs compared to \teff\
and \extinct. Yet differential weighting of these APs was used in the
fitness function (Table~\ref{param_table}) to make the optimization
more sensitive to these APs.  Clearly, these weak parameters still
present a problem for a 5-filter BBP system at G=15.

Mean AP-gradients up to 10--35\% higher (depending on the
AP) are found with the best filter systems from other runs. In other
words, filter systems which have a lower fitness may nonetheless
perform much better on a subset of the problem (e.g.\ for some APs or a
subset of sources).  This is almost inevitable when optimizing a
fitness function which is an aggregate of many separate
objectives. I shall return to this point in section~\ref{discussion}.

Fig.~\ref{fit260_7.sinalpha} shows the distribution over the six
orthovariance terms. Recall the transfer function used to give higher
weight to $\sin\alpha > 0.95$ (eqn. \ref{transfunc1}). HFD shows some
success in achieving this high degree of vector separation in five
of the six terms.  However, in these cases many sources are still
poorly separated, and as the mean values correspond to
$\alpha$=41\deg--45\deg, significant degeneracy clearly remains.
Furthermore, \teff\ and \extinct\ remain strongly degenerate for
almost all sources, with a mean angle of only 14\deg.\footnote{
Interestingly, if a filter system is optimized 
only on \extinct\ and \teff\ (but with the
grid unchanged), then the mean value of this $\sin\alpha$ is increased to 0.34.
This is also the only time that any sources
(ca.\ 40) are seen with $\sin\alpha > 0.95$.  Thus HFD does show some
ability to find filter systems which partially break the
\extinct,\teff\ degeneracy.}

To put this performance in context, and to assess the efficiency of
the HFD search and selection procedure, we look at the performance of
random filter systems. These achieve reasonable AP-gradient separation
for \teff\ and \extinct\ (caption to Fig.~\ref{fit260_7.apgrad}). This
is perhaps not that surprising given the broad band effects of APs
(Fig. \ref{sedplot}), because the filter systems are randomized between
the limits listed in Table~\ref{param_table} so will include many wide
filters. But the random systems perform somewhat worse on \feh\ and
\logg: the optimized systems increase these AP-gradients by 90\% and
50\% respectively. Likewise the six orthovariance terms in the optimized
the system are larger by factors of about 1.4 with respect to random
systems.

\subsubsection{Model parameter variations}\label{variations_bbp}

{\bf Number of filters}.
The baseline Gaia design calls for 4--6 filters in BBP on the grounds
of the chromatic correction for the astrometry (section~\ref{Gaia}).
A system of five filters was optimized above. While HFD can reduce the
number of effective filters (by assigning zero integration time), it
cannot add filters. The 20 runs were therefore repeated using 10
filters (and with the lower and upper limits on \itime\ set to 0.025
and 0.4 respectively). Of the resulting run maxima systems, 18 had 8--10
effective filters, although in several cases just a few filters
receive most integration time or there were some near-identical
filters. The other two filter systems had seven filters. These
were not only the fittest two systems, but they
also closely resemble the fittest filter systems from
the 5-filter optimization plus two additional filters with low
\itime. The fitnesses of these 20 run maxima lie in the range
22.3--24.8, i.e.\ slightly lower than the 5-filter optimization.
Inspection of the distributions of the fitness terms (cf.\
Figs.~\ref{fit260_7.apgrad} and~\ref{fit260_7.sinalpha}) for the two
7-filter systems shows that while they have lower mean AP-gradients
than the 5-filter systems, they have very slightly higher
orthovariance factors. This is even the case for some of the less fit
systems with more filters. Thus extra filters could contribute to
improved vector separation at the price of lower AP-gradients.

{\bf Magnitude}.
The end of section~\ref{fitness} discussed the dependence of the
fitness on the source magnitude.  Repeating the 5-filter optimization
at G=20 gives very different results: the seven fittest systems
consist of only two effective filters and the remaining 13 systems of
three.  The AP-gradients are much lower than expected from just
scaling from the G=15 results based on Poisson noise, indicating that
G=20 is already the faint star limit for some filters.  The systems
with fewer than four filters of course cannot determine four APs
independently. We may conjecture that these filter systems are
sensitive to fewer than four of the APs. However, this is not born out
by inspection of the distributions of the orthovariance factors, which
are all decreased by similar amounts ($>0.1$).  Clearly, HFD is unable
to find useful solutions when optimized on data at G=20.  However, if
the fitness terms for systems optimized at G=15 are recalculated at
G=20, then we find that the orthovariance terms are only reduced by a
few percent (see section~\ref{comparison}). Perhaps in the faint star
limit the fitness space is dominated by strongly attracting, poor
optima.
 
{\bf Grid}.
The HFD filter systems are of course very dependent on the stellar
grid. By way of illustration, if we restrict the grid to stars with
\teff\,$<$\,8000\,K, the optimized filter systems allocate more
integration time to the bluest filter, presumably to compensate for
the reduced flux in this wavelength range for the average star.  The
AP-gradients for \extinct, \logg\ and \feh\ are now 20--80\% higher
than those obtained with the full grid. (The AP-gradients for \teff\
are of course lower, because a given SNR-distance translates to a smaller
AP-gradient on account of the normalization of $\phi$ in
eqn.~\ref{AP-gradient}.)


{\bf Filter profile}.
Repeating the optimization with rectangular filter profiles increases
the typical run maximum fitness by about 5\%.  This is attributed
mostly to larger AP-gradients rather than to the orthovariance terms,
which show negligible difference.  The parameters of the optimized
filter systems are very similar to those obtained with the nominal
profile. The use of profiles with steeper sides therefore does not
help the vector separation (at the level of separation achieved here).

{\bf Elitism}. 
If elitism is not used ($E$=0), the evolution is very different
because the fittest systems are not forcibly retained.  The maximum
fitness now evolves in a similar erratic fashion to the minimum fitness
seen in Fig.~\ref{fit260_5.log}. Correspondingly, the evolution of the
filter system parameters shows no convergence (no `lines' as in
Fig.~\ref{fs260_5.log}), indicating a lack of selection pressure. The
run maximum is often found (and lost) within the first few tens of
generations.  These generally have a fitness around 15\% lower than
the run maxima attained when using elitism: it is clearly desirable to
force the retention of the best solutions to await favourable
offspring.
In contrast, if $E=100$ (half the population), the run maxima
fitnesses are more tightly bunched (24.9--26.1 compared to 23.4--26.1)
and there is a much higher degree of consistency across the
corresponding filter systems, although the maximum fitness across 20
runs is no higher.  Increasing the size of the elite is therefore
desirable, as it increases the reliability of the outcome and so
reduces the need to perform as many separate runs. Of course,
increasing $E$ beyond some point will be counterproductive as it
leaves fewer individuals available for search.

{\bf Population size}.
A sufficiently large population is required to maintain diversity. If
the population is too small it quickly becomes dominated by a
suboptimal filter system before there has been adequate opportunity
for search. Increasing the size of the population (and the elite) by a
factor of ten typically improves the run maximum fitness by only 4\%.

{\bf Recombination}.
If mutation is used as the only search operator, the resulting filter
systems are qualitatively unchanged, and the fitnesses (and mean
values of the individual fitness terms) are no lower.
This recombination operator is therefore redundant.  This is
not that surprising: randomly swapping a single filter between
filter systems is not obviously useful when we consider that it is the combined
effect of all filters which determines how well stars are separated.
This is in contrast to some other EA representations which use
crossover operators (see Appendix B) in which a parameter (gene) and
its local neighbours have a joint expression somewhat independently of
the other genes.

{\bf Mutation}.
HFD without mutation would not be useful, as only mutation creates new
filters. But the HFD results are not very sensitive to the probability
of mutation. If it is reduced by a factor of ten to 0.04 then the run
maximum fitness is changed by less than 1\%.  
Lowering the standard deviations of the mutations for all three
parameter types by a factor of two likewise has a negligible effect on
the final filter systems.


\subsection{Results: MBP}\label{results_mbp}

The main differences between the MBP and BBP instruments are shown in
Table~\ref{param_table}. MBP has a much larger integration time than
BBP and is intended for detailed astrophysical characterization.  A
systems of 10 filters is initially considered.

\begin{figure}
\centerline{
\psfig{figure=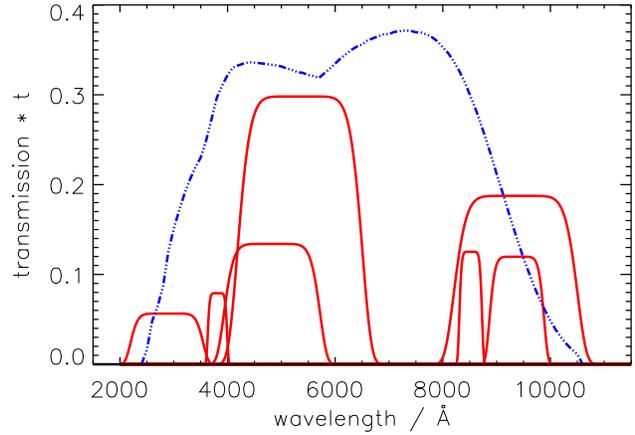,angle=90,width=0.50\textwidth}
}
\caption{The HFD-1M filter system, a MBP filter system produced by
HFD (solid lines). The CCD+instrument response is shown by the dot-dashed
line. See caption to Fig.~\ref{fs260_7.trans}.}
\label{fs300_9.trans}
\end{figure}

The fitness evolution for MBP is qualitatively the same as for BBP,
but now the run maximum fitnesses lie in the range 82--89, a factor of
more than three above BBP. The AP-gradients are considerably larger
than those found with BBP at the same magnitude, as expected due to
the larger sensitivity for MBP at this magnitude. The orthovariance
factors are similar to or larger than those found for BBP.  The
fittest filter system from a set of 20 runs is shown in
Fig.~\ref{fs300_9.trans}. It shows a mix of broad and narrow band
filters extending to the most extreme wavelengths permitted by the
instrument/CCD response.  There are some similarities to the BBP
systems shown in Fig.~\ref{fs260_runs.trans}, e.g.\ the relatively
narrow filter around 8500\,\AA\ (see Table~\ref{hfdfs_params}).

\begin{table}
\begin{center}
\caption{Parameters for the filters in the HFD-1B (BBP) and HFD-1M
(MBP) filter systems according to the filter parametrization given in
section~\ref{representation}. Wavelengths are reported to $\pm 5$\,\AA.
In a practical implementation, filters with profiles extending beyond
the CCD QE cutoff would be truncated, thus altering \cw\ and \hwidth.}
\label{hfdfs_params}
\begin{tabular}{llrrr}
~\\
\hline
         &      & \cw\ / \AA\ & \hwidth\ / \AA\ & \itime\ \\
\hline
HFD-1B   &  1   & 4545 & 1505 & 0.194 \\
         &  2   & 6125 &  555 & 0.133 \\ 
         &  3   & 7470 &  490 & 0.336 \\
         &  4   & 8560 &  115 & 0.208 \\
         &  5   & 9440 &  675 & 0.129 \\
\hline
HFD-1M   &  1   & 2870 &  660 & 0.056 \\
         &  2   & 3810 &  175 & 0.079 \\
         &  3   & 4825 &  870 & 0.134 \\
         &  4   & 5345 & 1130 & 0.298 \\
         &  5   & 8520 &  205 & 0.125 \\
         &  6   & 9355 & 1120 & 0.187 \\
         &  7   & 9375 &  495 & 0.120 \\
\hline
\end{tabular}
\end{center}
\end{table}

One may postulate that higher orthovariance factors could be achieved
with narrower filters, as these could plausibly better discriminate
between spectral features. This was tested by repeating the
optimization with the maximum HWHM set to 400\,\AA.  The result is
that the orthovariance factors are increased by about 0.05 (more for
\extinct,\teff), although not many more are put into the desired
0.95--1.00 range required to reasonably break the degeneracy. This
also comes at the cost of reduced AP-gradients -- as fewer photons are
collected -- although these are still adequate. Interestingly, the
optimization does not force all the filter widths to the maximum
permitted. The fitness is about 20\% lower.  The improvement in vector
separation from narrow filters is therefore small 
(and the degradation in AP-gradients may not be acceptable at faint
magnitudes).  This is not surprising when we again
consider that the effects of AP variations are coherent over a wide
wavelength range (Fig.~\ref{sedplot}).

 Many of the HFD-optimized MBP systems include filters with only a
small allocation of fractional integration time, \itime. It is
therefore tempting to think that such filters could be removed and
their integration time allocated to other filters, but in fact this
frequently results in a dramatic decrease in the fitness. (This is the
case for the HFD-1M system, for example). 

The optimization leading to HFD-1M was done with 10 filters, but there
are only 7 effective filters in this optimized system. Across the set
of 20 runs, two had 7 effective filters, one had 8 and the rest 9 or
10, although those with 9 or 10 filters often have two or more very
similar or extensively overlapped filters.  If the optimization is
repeated with 15 nominal filters, the run maxima fitnesses and
orthovariance terms are similar or slightly lower than with 10 nominal
filters. With only 5 nominal filters in the MBP optimization,
the resulting systems sometimes have slightly higher fitness
than the nominal 10-filter MBP systems. However, they have smaller
orthovariance factors, which, given that the 10-filter systems already
achieve adequate AP-gradients, is more significant.

In conclusion: at the level of separation currently achieved, 7 or 8
filters in MBP are an optimal trade off between spectral sampling and
sensitivity for the grid in Table~\ref{grid_table}.

\subsection{Comparison with other photometric systems proposed for Gaia}\label{comparison}

\begin{figure}
\centerline{
\psfig{figure=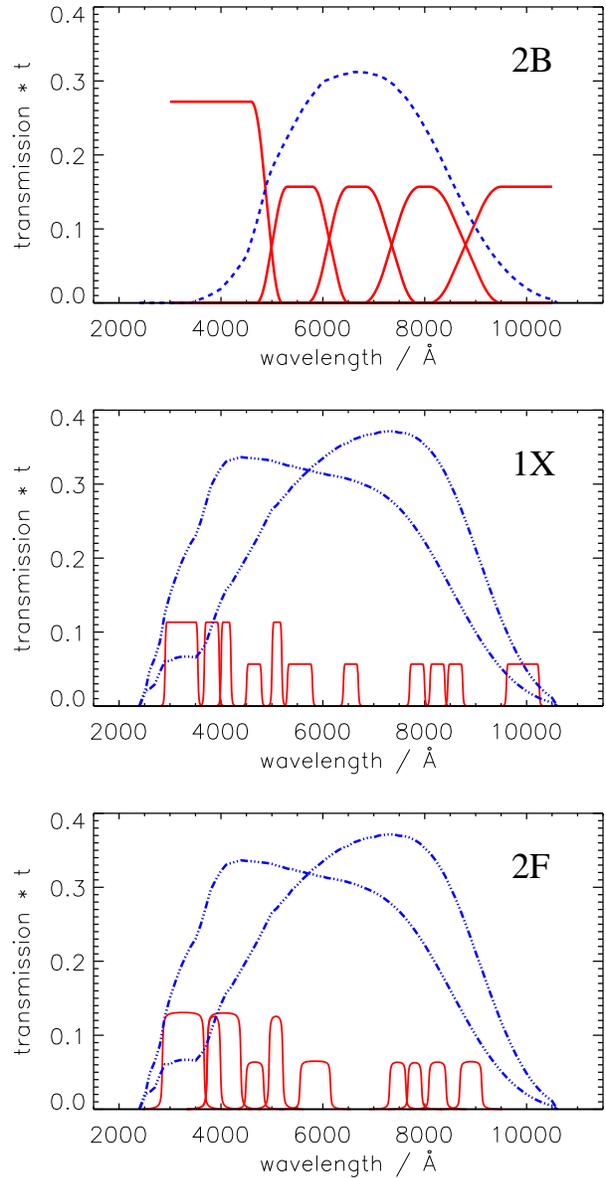,angle=0,width=0.50\textwidth}
}
\caption{Three filter systems proposed for Gaia by previous authors.
The filter profiles have been plotted in the same way as in
Fig.~\ref{fs260_7.trans} to show the fractional integration
times. Over plotted in each case are the CCD sensitivities.  2B is a
BBP system; 1X and 2F are MBP systems.  }
\label{otherfs}
\end{figure}

The HFD fitness function is a general figure-of-merit and can be
calculated for any photometric system. A number of other photometric
systems have been designed for Gaia. The present main candidates are
the BBP 2B system (Lindegren \cite{lindegren03}) consisting of 5
broad, partially overlapping filters and, for MBP, 2F (Jordi et al.\
\cite{jordi03}) and 1X (Vansevicius \& Bridzius \cite{vansevicius03}),
both consisting of relatively narrow filters to measure specific
stellar features (Fig.~\ref{otherfs}). The fitness terms have been
calculated for these filter systems using exactly the same instrument
models and grid as used for the HFD optimization.
Table~\ref{comparison_table} lists the fitness terms and compares them
with the HFD systems.

\begin{table}
\begin{center}
\caption{Fitness comparison of the two HFD filter systems HFD-1B and
HFD-1M and three other systems proposed for Gaia (Fig.~\ref{otherfs}).
The overall fitness as well as the mean values (averaged over the
sources in the grid) of the AP-gradients and the orthovariance terms
are shown. The quantities are calculated with all sources at G=15.}
\label{comparison_table}
\begin{tabular}{lrrrrr}
\\
\hline
                             & HFD-1B  & 2B   & HFD-1M  & 1X   & 2F   \\
\hline
Fitness                      & 26.1 & 13.7 & 89.4 & 34.6 & 34.2 \\
h(\extinct)                  & 1832 & 1944 & 7206 & 3183 & 3932 \\
h(\feh)                      &   94 &   72 &  272 &  195 &  226 \\
h(\logg)                     &  103 &   86 &  337 &  240 &  261 \\
h(\teff)                     & 1765 & 1888 & 6827 & 3177 & 4095 \\
$\sin\alpha$(\extinct,\feh)  & 0.70 & 0.66 & 0.70 & 0.76 & 0.73 \\
$\sin\alpha$(\extinct,\logg) & 0.68 & 0.60 & 0.75 & 0.81 & 0.77 \\
$\sin\alpha$(\extinct,\teff) & 0.24 & 0.24 & 0.21 & 0.40 & 0.40 \\
$\sin\alpha$(\feh,\logg)     & 0.71 & 0.61 & 0.73 & 0.73 & 0.71 \\
$\sin\alpha$(\feh,\teff)     & 0.66 & 0.63 & 0.68 & 0.70 & 0.66 \\
$\sin\alpha$(\logg,\teff)    & 0.66 & 0.60 & 0.75 & 0.77 & 0.73 \\
\hline
\end{tabular}
\end{center}
\end{table}

For BBP, HFD-1B has almost twice the fitness as 2B although only two
of its AP-gradients are higher. Its higher fitness is due to the fact
that it has quite a few more sources with orthovariance terms in the
desired range 0.95--1.00, which is given more weight in the fitness
due to the transfer function (eqn.~\ref{transfunc1}).  This difference
between the distributions is not represented by the only slightly
higher (unweighted) mean values in Table~\ref{comparison_table} for
the orthovariance terms for HFD-1B.

Turning to MBP, HFD-1M is 2.6 times fitter than either 1X or
2F. HFD-1M achieves much higher AP-gradients by virtue of its wider
filters. 
It has slightly lower mean orthovariance terms than does 1X.  This
agrees with what was observed with HFD systems
(section~\ref{results_mbp}), namely that narrower filters can achieve
better vector separation, although in both cases the improvement is
small.  More importantly, 1X has only slightly more sources with
orthovariance terms in the range 0.95--1.00 than does HFD-1M (and the
latter actually has more values of $\sin\alpha$(\logg,\teff) in this
range).  As this is the range we are primarily interested in (as only
then can we say that degeneracy has been satisfactorily minimized), we
see that 1X is little better than HFD-1M at vector separation. Only
for \extinct,\teff\ does 1X achieve a much higher mean, although it
too has no sources in the 0.95--1.00 range. Thus 1X -- like HFD-1M --
has failed to break the degeneracy between extinction and effective
temperature.  Comparing HFD-1M with 2F, the latter has slightly fewer
sources in the 0.95--1.00 range than the former so is 
slightly worse at vector separation. 

If the fitness terms are recalculated at G=20, then the fitnesses of
all filter systems are reduced by a similar factor. The AP-gradients
of course decrease a lot as they are just proportional to the SNR.
For HFD-1B the mean values are 60, 3, 3, 57 (order as in
Table~\ref{comparison_table}); for HFD-1M they are 281, 10, 13 and
263. The values are correspondingly lower for the other filter
systems.  For BBP these all fall well below the desired value of
around 200 (see section~\ref{nominal_bbp}), so good scalar separation
is not possible for many stars with BBP at the limiting magnitude of
the survey.  For MBP, good scalar separation is possible at G=20 for
extinction and effective temperature, but not for metallicity or
surface gravity.  (This should only be taken as a rough indication,
however, because in parts of the grid AP-gradients much less than
200 are acceptable, e.g.\ for \feh\ in hot stars.) At G=20 the
orthovariance factors are only decreased by a few percent with respect
to G=15. This is what we would expect as the principal directions are
little affected by the SNR.

In summary, we find that the two HFD systems perform as well as or
better than their `classical' (2F, 2B, 1X) counterparts.  Nonetheless,
there are still some poor aspects of the HFD (and classical) filter
systems, the possible cause of which will be discussed in the
following section. Once these have been improved upon, a more detailed
assessment of the parametrization performance of the HFD systems using
standard methods (e.g.\ Bailer-Jones \cite{bj00}) will be warranted.

\section{Discussion: HFD assumptions and improvements}\label{discussion}

A critical analysis of HFD follows to highlight the underlying
assumptions and weaknesses of the approach, along with suggestions of
how it may be improved.

HFD as implemented in this article is concerned only with determining
stellar parameters. In a survey, the same filter system must also
distinguish single stars from the `contaminants', such as quasars,
unresolved galaxies and unresolved binary stars. (With Gaia, some
assistance in this task comes from the astrometry.)  The filter system
could be simultaneously optimized to distinguish such contaminants by
adding an extra term to the fitness which is the sum of SNR-distances
between each contaminant and each source in the grid. Maximizing this
places the contaminants away from the sources of interest.

The filter systems produced by HFD depend on the grid of APs, the SEDs
used to represent the stars and the weights set for each AP. As the
fitness is just a sum over all sources, the relative distribution of
different types of stars is significant and must be carefully
considered. Note also that the fitness sum (eqn.\
\ref{nominalfitness}) may be generalized to include different weights
for each star or even for each AP of each star.

It should be emphasized that the present fitness function is concerned
only with a {\it local linear} separation of sources in the
multidimensional filter space. If successful, it means that only
locally linear regression models are necessary to calibrate this data
space in terms of APs, allowing considerable simplifications.  Higher
order terms could be included in the fitness function and generally
one would expect this to give rise to superior filter systems -- or at
least a more accurate determination of the fitness -- at the cost of a
more complex optimization.  Calibration would correspondingly require
locally nonlinear regression methods.  A further complication which is
ignored in HFD is the possible presence of {\it global} degeneracies
of APs, i.e.\ disjoint parts of the AP space overlapping in the data
space.  Ideally the fitness function should be further modified to
measure the extent of this and penalize against it.

The grid used (Table~\ref{grid_table}) is relatively sparse in the
APs.  This is probably adequate from the point of view of sampling the
dependence of the data on the APs.  However, this same grid is used to
provide the neighbours (the isovars; see Fig.~\ref{hfd_dist2}) for
determining the fitness at each grid point. HFD therefore implicitly
assumes that the photon counts vary linearly with the APs on the local
scale between a source and its isovars. This assumption may not be
valid for all points in the grid. It could be avoided by using a
second, denser grid from which the isovars are selected to better
satisfy the local linearity assumption.\footnote{Needless to say,
nothing is gained if this second grid is constructed by linearly
interpolating the first grid.}

One of the major limitations of the HFD filter systems (and the others
considered in section~\ref{comparison}) is that they give relatively
poor vector separation for many stars, i.e.\ APs remain degenerate in
parts of the grid.  It remains to be analysed in
detail whether these are regions of the grid which are intrinsically
degenerate for medium and broad band photometry, or whether HFD is
simply unable to create suitable filters systems.  As
the fitness function is an amalgamation of different terms (4
AP-gradients and 6 orthovariances), there is a danger that a high
fitness be achieved by increasing some terms with little regard to the
latter, i.e.\ the optimization becomes desensitized to some of the
fitness terms.  This will be returned to below. This is likewise a
problem for different sources: a high fitness could be achieved by
overseparating some sources while underseparating others.

An alternative explanation for poor vector separation is that the
search operators are not searching the parameter space efficiently.
An efficient, directed (rather than random) search is important
because there are a very large number of potential filter systems, as
a simple calculation makes clear. Suppose that only differences in
central wavelength by at least 100\,\AA\ are significant.  In this
case, the wavelength range 4000--10\,000\,\AA\ contains only 60
different central wavelengths.  Applying the same step size to HWHM
increments of up to 1500\,\AA\ gives 15 discrete filter widths.  Even
if we ignore variation in the fractional integration
times,
for a system of 5 filters there are of order $10^{14}$
different combinations of filter parameters and of order $10^{29}$ for
systems with 10 filters, of which only a negligible fraction can be
ignored as being obviously inappropriate.  In contrast, HFD evaluates
around $10^5$ filter systems during an optimization run.

Provided the offspring of fitter parents are generally fitter than the
offspring of less fit parents (which has been experimentally confirmed
with HFD), the population will evolve toward fitter solutions
(Hinterding \cite{hinterding00}).  Elitism then guarantees that the
optimum is found (eventually). If the fitness convergence seen in
Fig.~\ref{fit260_5.log} is asymptotic, then the search operators are
working well, i.e.\ the solutions we find are about the fittest
available. However, it could be that some specific changes of the
filter system parameters produce a significant increase in fitness.
If this is the case, then more directed search operators may be
useful. One possibility is to use a hybrid stochastic/gradient search
method.  Another is to use {\it strategy parameters} to adapt the size
of the mutations as the evolution proceeds (see Appendix B).  I
attempted a variation of this which reduced the mutation sizes for a
single filter system if its fitness decreased -- the rationale being
that as an optimum is approached a more refined search should be
undertaken -- but this did not lead to improvement.


These considerations aside, my feeling is that the most significant
limitation of HFD is the fact that it consists of only a single
objective function.  It cannot be overemphasized that our goal is
really to simultaneously maximize ten separate fitness terms: 4
AP-gradient and 6 orthovariance terms (each averaged over the sources
in the grid).  Eqn.s~\ref{vecsep} and~\ref{fitmeas} is but one way to
amalgamate these into a single objective function, albeit with some
justification.  Nonetheless, it can result in less fit filter systems
yielding higher values for some fitness terms than do fitter filter
systems, as was seen earlier. For example, the optimum from run 19 in
Fig.~\ref{fs260_runs.trans} achieves AP-gradients 5--20\% higher than
the fittest filter system (run 7 in the same figure), yet its fitness
is 4\% smaller due to lower orthovariance factors.  How can we
properly compare two filter systems which have very similar overall
fitnesses, yet one performs better at some aspects and worse at
others?  In principle, the fitness function is suitably constructed
and weighted to increase monotonically with increasingly `better'
filter systems.  But can we uniquely establish in advance what
`better' means?  Not only is it very difficult to determine an
appropriate weighting a priori, it is strictly not possible. This is
because (a) we cannot compare different types of measures (e.g.\
AP-gradients and orthovariance terms) on an equal footing, (b) the
scientific criteria may not give rise to a unique weighting, yet the
optimum solution may be quite sensitive to this weighting, and (c) any
attempt to weight terms a priori requires that we have some idea of
what degree of separation is even possible within the constraints of
the instrument and grid, yet this is something we generally do not
know in advance. Thus a single fitness function could easily set
conflicting or unattainable requirements on the different terms.

A solution to this dilemma is available through the use of
multiobjective optimization methods (e.g.\ Deb \cite{deb01}). This
approach avoids comparing dissimilar objective functions by optimizing
each separately.  The goal is not to arrive at a single solution, but
at a set of so-called `non-dominated solutions'.  A non-dominated
solution is one for which no other solution exists which has higher
values of all objective functions.  The non-dominated set of solutions
over the entire search space is called the `Pareto optimal set',
solutions which are better than all other possible solutions in all
objectives, but are only better than each other in some respects. They
are, therefore, the best possible set of compromise solutions.  Having
found these we can reassess the scientific requirements in terms of
what is actually possible within the optimization constraints and
select the most desirable compromise.

\section{Summary and conclusions}\label{conclusions}

I have introduced a novel approach to the design of photometric
systems via optimization of a figure-of-merit of filter system
performance.  In the present incarnation, this figure-of-merit (or
fitness) measures the ability of a filter system to determine multiple
stellar astrophysical parameters (APs), by calculating the separation
in the data (filter) space between stars with different APs.  The
better that sources can be separated (in signal-to-noise units) according
to their AP differences, the better the filter system.  This
separation is vectorial in nature, meaning that the figure-of-merit is
also proportional to the angle between the vectors which define
the directions of local variance of each AP: In the ideal filter
system these vectors would be mutually orthogonal at all points in the
AP space, thereby removing any degeneracy between APs. The fitness is
calculated via an instrument model for a grid of spectra, which
sample stellar parameters the photometric system must
determine.

The optimization is performed with an evolutionary algorithm. In this
approach, a population of filter systems is evolved according to the
principle of natural selection, such that the fitter filter systems
are more likely to survive and to produce more `offspring'.
Reproduction takes place by combining or mutating selected parents,
resulting in changes of the filter parameters (central wavelength,
profile width, integration time), thus providing a stochastic yet
directed search of the filter parameter space.

This model, HFD (Heuristic Filter Design), has been applied to design
CCD photometric systems for the Gaia Galactic Survey Mission.  The
systems were optimized to separate the four APs effective temperature,
\teff, metallicity, \feh, surface gravity, \logg, and interstellar
extinction toward the star, \extinct.  Recurrent characteristics of
the resulting filter systems are broad overlapping filters, although
filters with a half-width above 1500\,\AA\ were consistently
disfavoured.  The preferred broadness is not surprising when one
realises that each of the APs has a coherent effect on the data over a
wide wavelength range.  Narrower filters were found not
to improve significantly the orthogonality (vector separation). This
tendency toward broader filters than have hitherto been adopted for
the Gaia filter systems -- and for stellar parametrization in general
-- is one of the main results of this application of HFD. Likewise is the
related tendency toward overlapping filters.  This may be indicative
of a more efficient use of a multi-dimensional data space than
non-overlapping systems.

In terms of the scalar separation of sources, the HFD filter systems
perform well at the Gaia target magnitude of G=15, although at the
limiting magnitude of G=20 the separation for \feh\ and \logg\ is
unsatisfactory. More significantly, the vector separation is
inadequate in parts of the AP grid, and between \teff\ and \extinct\
in particular considerable degeneracy remains. Yet other systems
proposed for Gaia show similar difficulties, and overall HFD performs
at least as well as or better than these.  It remains to be seen whether
these are intrinsic limitations of broad and medium band photometry
for these instrument models of whether improvements to the fitness
function alter this.  Either way, this systematic approach to filter
system design embodied in HFD shows considerable promise.

A number of improvements to HFD to address some deficiencies were
suggested, including the use of more efficient search operators, the
use of secondary grids or generalization to nonlinear separation, and
the incorporation of multiobjective optimization methods. The latter
allows the different objectives of the filter system to be optimized
separately, thus avoiding having the problem of weighting and combining
heterogeneous objectives.

Specifically with regard to Gaia, HFD may be developed in a
number of ways. The most significant is perhaps the inclusion of
parallax information: the parallaxes from Gaia will permit an accurate
determination of the luminosity and (via \teff) the radius of many
stars, reducing the need to determine \logg. By including the parallax
error model in HFD, filter systems better matched to the available
astrometry can be designed.

Beyond this application, HFD represents a generic approach to
formalizing filter design by casting it as an optimization problem
with few prior assumptions. The key steps are the parametrization of
the filter system, the construction of a figure-of-merit, and the
design of appropriate genetic operators to search the parameter space.
Evolutionary algorithms are particularly appropriate for this problem
because the fitness landscape in which the optimization is performed
will is frequently complex and noisy.  While these steps are
nontrivial, HFD provides a general framework for applying this
approach to many other problems. These include the identification of
particular types of objects, such as ultra cool dwarfs or metal poor
stars, star/quasar separation, the spectral classification of
galaxies, and photometric redshift determination.  This is applicable
not only to future large scale surveys, but also for more modest
surveys on existing ground-based facilities.

\section*{Acknowledgements}

I would like to thank Anthony Brown, Jos de Bruijne, Lennart Lindegren
and Michael Perryman for useful suggestions and discussions during the
development of HFD. I particularly acknowledge input and constructive
criticism from Jos de Bruijne in the early stages of this work.
Thanks also to Carme Jordi for information concerning the instrument
model.  I am especially grateful to Bob Nichol and the Astrostats
group at Carnegie Mellon University for access to significant amounts
of computer time and for general support during my time in Pittsburgh.


\section*{Appendix A: Fitness functions}

\begin{figure}
\centerline{
\psfig{figure=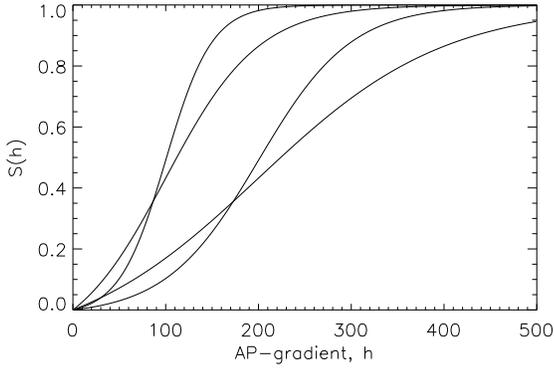,angle=90,width=0.40\textwidth}
}
\caption{A transfer function for the AP-gradients, given by
eqn.~\ref{tf2_eq}. From left to right at the top of the plot, the
lines are for $\kappa$=100,$\omega$=2, $\kappa$=100,$\omega$=4, and $\kappa$=200,$\omega$=2, $\kappa$=200,$\omega$=4.}
\label{tf2_fig}
\end{figure}

HFD uses single objective optimization, requiring the aggregation of
the AP-gradients and the orthovariance terms into a single objective
function.  This was achieved via the cross product
(eqn.~\ref{crossproduct}).  However, it was found advantageous to modify the
basic form. First, the AP-gradients were raised to a fractional power
(1/2) to avoid overseparation (a subset of sources dominating
the fitness).  Second, the four APs were assigned different weights to
account for the fact that each has a different magnitude effect on the
data. These two problems could in principle we avoided by using a
transfer function to map the full range of the AP-gradient (zero to
essentially infinity) to a restricted common range for all APs, say
0--1. One possible form is a modified sigmoid function
\begin{equation}
S(h) = e^{-\omega} \left( \frac{ 1+e^\omega }{ 1 + e^{-\omega(\frac{h}{\kappa}-1)} } - 1 \right)
\label{tf2_eq}
\end{equation}
shown in Fig.~\ref{tf2_fig}. $S(h)$ replaces $w_j h$ in
eqn.~\ref{fitmeas}. $\kappa$ is the
transfer point of the function and $\omega$ controls the sharpness of
the transfer.  As discussed in section~\ref{fitness}, $\kappa=200$ is
a reasonable target value for AP-gradients.  Values of $h$ well below
this are assigned low fitness, and all values well above $\kappa$ are
assigned a similar fitness, thus avoiding overseparation. (With the
AP-gradients now mapped onto the same range as the orthovariance
terms, we could even avoid the cross product approach by simply
summing the terms.)

The major problem with a transfer function is that it disregards what
values of the APs it is even possible to achieve with the instrument
model and grid. That is, simply assiging $\kappa$ to achieve target
values of $h$ for some AP may force $S(h)$ into the unresponsive
regions near 0 or 1, making all sources equally fit or unfit, thus
depriving the fitness function of discriminative power.  With
$\kappa=200$, $S(h)$ would be near 0 for \feh\ and \logg\ for almost
all sources, whereas it would be near 1 for \extinct\ and \teff\ for
almost all sources (see Fig.~\ref{fit260_7.apgrad}). Thus not only
would the fitness be indiscriminative, it would also be insensitive to
\feh\ and \logg.  Experimental testing of this transfer
function confirms this. We cannot assign target values according to
ideals to arrive at a useful fitness measure.

In contrast, the static weighting used in eqn.~\ref{fitmeas}
brings all of the APs to the same `level' in the fitness function,
forcing the fitness and hence the selection operator to be equally
sensitive to all APs. However, these static weights cannot be
determined a priori as they depend not only on the grid and instrument
model, but also on the typical AP-gradients and hence the filter
system itself, thus demanding an unsatisfactory quasi-iterative approach.

The underlying problem of all of these approaches arises from the need
to combine different objectives (AP-gradients, orthovariance terms)
into a single fitness function which both takes account of the
different effects they have on the data and ensures that they are
equally represented.  This is rather intractable, as the effects of
the APs on the data depend on the filter system itself, just the thing
we are trying to modify using the fitness. The way around this is not
to combine the different objective functions at all, but to optimize
each separately using multiobjective optimization methods (see
section~\ref{discussion}). This would obviate several unsatisfactory 
aspects of the HFD model and the need for the above transfer function.

\section*{Appendix B: Evolutionary algorithms}

There are many variants on how the genetic operators (selection,
recombination and mutation) are implemented and, equally importantly,
how the problem parameters (genes) are represented. Historically, the
approaches can be split into at least three broad (and overlapping)
categories: {\it genetic algorithms}, {\it evolution strategies} and
{\it evolutionary programming}.

Genetic algorithms (GAs) typically use a binary representation for the
genes, that is, each individual is represented by a string of
binary digits.
Recombination usually involves probabilistically selecting
two individuals from the parent population, randomly choosing a point
between two genes at which both individuals are split, and then
recombining the left part of one with the right part of the other and
vice versa, to create two new individuals. This is so-called `single
point crossover'. Repeating this for $\mu/2$ randomly selected pairs
for a population size $\mu$ produces a new population. Mutation takes
a comparatively background role, randomly flipping one or more genes
with low probability to avoid stagnation.

Evolution strategies (ESs), on the other hand, usually use
real-valued representations, i.e.\ $K$ real-valued genes.
Recombination may be used, but mutation is often a more significant
operator: it is applied to all $\mu$ individuals in producing the
intermediate population, usually by adding a Gaussian random variable
N$(0,\sigma_k)$ to the $k^{th}$ gene.  Another feature of the
canonical ES is that these mutation parameters, $\{\sigma_k\}$ (the
so-called {\it strategy parameters}), are themselves mutated at each
generation, a procedure referred to as {\it self adaptation}. In other
words, the typical mutation sizes are themselves subject to natural
selection. This may even be extended to include covariance strategy
parameters.  Selection is usually deterministic with ESs, either using
$(\mu,\lambda)$ selection, in which an offspring population of size
$\lambda$ is produced and the $\mu$ fittest individuals are selected,
or $(\mu + \lambda)$ selection, in which the $\mu$ fittest individuals
are selected from the union of the $\lambda$ offspring and the $\mu$
parents. These are highly elitist strategies.

Evolutionary Programming (EP) is closely related to ESs, the two main
differences being the employment of probabilistic selection and the
exclusive use of mutation in EPs.

This broad distinction into ESs, GAs and EPs is largely historical
and many applications now draw upon elements of each.  HFD is no
exception in this respect.

Since the introduction of what are now collectively called
evolutionary algorithms for optimization in the 1950s and 1960s, they
have undergone considerable development and have been applied in a
variety of fields.  There is a vast literature on GAs, ESs and EPs. An
introduction to all three types of evolutionary algorithm can be found
in B\"ack \& Schwefel (\cite{baeck93}) or Fogel (\cite{fogel95}), with
more comprehensive information on many aspects provided by the
collection of articles edited by B\"ack, Fogel \& Michalewicz
(\cite{baeck00a}, \cite{baeck00b}).  A good introduction to GAs in a
broad sense is Mitchell (\cite{mitchell96}) and one of the classic,
most cited works in this field is Goldberg (\cite{goldberg89}).  A
discussion of the analogies employed in GAs (and simulated annealing)
can be found in Bailer-Jones \& Bailer-Jones (\cite{bjbj02}).

\end{document}